\documentclass[%
 reprint,
 amsmath,amssymb,
 aps,
pre,
superscriptaddress
]{revtex4-2}
\usepackage{subcaption}
\usepackage{caption}
\usepackage{ragged2e}
\usepackage{float}
\usepackage{booktabs}
\usepackage{graphicx}
\usepackage{dcolumn}
\usepackage{bm}
\usepackage[mathlines]{lineno}
\usepackage{enumitem}
\usepackage{makecell}
\usepackage{array}
\usepackage{amsfonts}
\usepackage{amsmath}
\usepackage{amsthm}
\usepackage{epsfig}
\usepackage{graphicx}
\usepackage[usenames,dvipsnames]{color}
\usepackage[utf8]{inputenc}
\usepackage{booktabs}
\usepackage{comment}
\usepackage[normalem]{ulem}
\usepackage{cancel}

 \usepackage[hidelinks,unicode=true]{hyperref}
 \hypersetup{colorlinks=true,
 	linkcolor=blue,
 	urlcolor=blue,
 	citecolor=blue,
 	pdfhighlight=/N
 }

\newcommand{\orcid}[1]{\hspace{0.2em}\href{https://orcid.org/#1}{\includegraphics[keepaspectratio,width=0.7em]{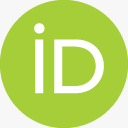}}}

\begin{document}

\title{Efficient Gillespie algorithms for spreading phenomena in large and heterogeneous higher-order networks}

\author{Hugo P. Maia\orcid{0000-0002-8912-0900}}
\affiliation{Departamento de F\'{\i}sica, Universidade Federal de Vi\c{c}osa, 36570-900, Vi\c{c}osa, MG, Brazil}

\author{Wesley Cota\orcid{0000-0002-8582-1531}}
\affiliation{Departamento de F\'{\i}sica, Universidade Federal de Vi\c{c}osa, 36570-900, Vi\c{c}osa, MG, Brazil}

\author{Yamir Moreno\orcid{0000-0002-0895-1893}}
\affiliation{Institute for Biocomputation and Physics of Complex Systems (BIFI), University of Zaragoza, Zaragoza 50018, Spain}
\affiliation{Department of Theoretical Physics, Faculty of Sciences, University of Zaragoza, Zaragoza 50009, Spain}

\author{Silvio C. Ferreira\orcid{0000-0001-7159-2769}$^{*}$}
\thanks{Corresponding author.}
\email{silviojr@ufv.br}
\affiliation{Departamento de F\'{\i}sica, Universidade Federal de Vi\c{c}osa, 36570-900, Vi\c{c}osa, MG, Brazil}
\affiliation{National Institute of Science and Technology for Complex Systems, Centro Brasileiro de Pesquisas F\'{\i}sicas, Rua Xavier Sigaud 150, 22290-180, Rio de Janeiro, Brazil}
\affiliation{Instituto de Ci\^{e}ncias Matem\'{a}ticas e de Computa\c{c}\~{a}o, Universidade de S\~{a}o Paulo, S\~{a}o Carlos, SP 13566-590, Brazil}


\begin{abstract}

Higher-order interactions, where groups of nodes interact collectively rather than pairwisely, are central to many complex systems, from neural and ecological networks to social contagion. However, simulating dynamical processes on such higher-order structures remains computationally challenging due to the combinatorial growth of possible interactions. Here, we develop efficient and statistically exact Gillespie algorithms for Markovian spreading dynamics on large and heterogeneous hypergraphs. By incorporating phantom processes $-$events that advance time without altering the system’s state$-$, we drastically reduce the computational complexity of standard algorithms ($\mathcal{O}(N^2)$), achieving up to linear scaling with system size. Relying on the susceptible–infected–susceptible model with critical mass thresholds as a benchmark, we show that the optimized algorithms outperform standard approaches by several orders of magnitude, enabling simulations of networks with millions of nodes and broad heterogeneity in both degree and interaction order. Efficient sampling methods, needed to overcome the bottlenecks imposed by either a high maximum order or number of interactions, and other dynamical processes on higher-order networks are tackled. These results establish a general framework for scalable, continuous-time simulations of higher-order contagion and related dynamical processes.

\end{abstract}

\maketitle


\section*{\label{sec:Intro}Introduction}

Networks provide the fundamental substrate for understanding how collective phenomena emerge from interactions among many agents~\cite{Barrat,Barabasi}. Classical network theory assumes that interactions are exclusively pairwise, an assumption that has guided decades of research on synchronization~\cite{Gomez-Gardenes2011,Arenas2008}, opinion formation~\cite{Deffuant2000,Maia2021}, brain activity~\cite{Haimovici,Bullmore}, and ecological stability~\cite{Montoya,Fath}, as well as on the spreading of diseases or information across populations~\cite{Pastor-Satorras2015,DeArruda2018PR}. Yet, in many real systems, interactions are inherently collective and cannot be reduced to independent pairwise contributions. In such cases, the dynamics depend on simultaneous, group-level processes that involve more than two units $-$phenomena that are naturally described within the framework of higher-order networks~\cite{Bianconi2021,Battiston2021}.

Simple networks describe systems whose elements interact pairwisely. In such cases, collective behaviors emerge from links connecting two agents at a time. While nonlinear effects of multiple pairwise interactions can sometimes be mimicked by effective rules, genuine group interactions cannot be decomposed into independent pairs~\cite{Battiston2021}. The natural extension of pairwise networks is given by higher-order networks~\cite{Bianconi2021}, which allow simultaneous interactions among any number of nodes and thus capture complex collective effects that simple graphs cannot represent. In pairwise network dynamics, the infinitesimal generator of the Markov process is decomposable into a sum of dyadic operators such that higher-order terms appear only at the level of observables~\cite{Hinrichsen2000} (e.g., correlation functions or closures). By contrast, in higher-order interactions represented by hypergraphs~\cite{Bianconi2021}, the microscopic transition rate is defined directly on groups of size $m>2$; the generator itself contains irreducible multi-body operators. Such rates cannot be written as a superposition of pairwise contributions without altering the dynamics.

Higher-order interactions have been shown to alter the behavior of complex systems fundamentally. They can induce abrupt or hybrid phase transitions, multistability, or hysteresis~\cite{deArruda2023}, phenomena that rarely have counterparts in pairwise models. Empirical evidence for such collective mechanisms spans diverse domains: synchronous activation of neural assemblies~\cite{Sizemore1,Sizemore2}, multi-species dependencies in ecological webs~\cite{Bairey}, and contagion events driven by group gatherings, airborne exposure, or social reinforcement~\cite{Mancastroppa,deArruda2024,St-Onge2021}. Capturing these effects requires frameworks where interactions among sets of nodes$-$represented as hyperedges or simplices$-$ govern the dynamics.

Despite recent theoretical advances, simulating dynamical processes on higher-order networks remains a major computational challenge. The number of potential interaction events grows combinatorially with system size and interaction order, leading to prohibitive computational costs even for moderately large systems. As a result, most previous studies have been limited to small networks or to low-order interactions (e.g., triads), often implemented in discrete time~\cite{Iacopini,Landry,Kim}. However, discrete-time approximations can introduce biases in continuous-time Markovian dynamics~\cite{Fennell,Silva2024}, and many key collective effects $-$such as vanishing epidemic thresholds in heterogeneous networks$-$, only emerge in sufficiently large systems~\cite{Pastor-Satorras2015,Ferreira2012}.

Here, we develop efficient, statistically exact algorithms for simulating continuous-time contagion dynamics on higher-order networks of arbitrary size and heterogeneity. We develop algorithms to account for highly complex combinatorial states of hyperedge activation, building on the concept of phantom processes~\cite{Cota2017} $-$ attempted events that advance simulation time without changing the system state $-$ which allows for drastic optimizations of higher-order contagion dynamics. Existing exact stochastic methods scale poorly in this setting because potential events cannot be enumerated locally without repeatedly scanning high-order structures.

This limitation reflects a fundamental difference with respect to pairwise dynamics. In standard network-based models, transition rates can be decomposed into contributions associated with individual nodes or edges, so that the infinitesimal generator of the process is a sum of local operators. As a consequence, the set of feasible events is locally enumerable and can be updated through strictly local operations after each microscopic transition. In contrast, the higher-order dynamics studied here involve intrinsic multi-body interaction rules, where the activation of a transition depends on the simultaneous state of multiple nodes within a hyperedge. This implies that the generator cannot, in general, be expressed as a superposition of pairwise contributions, and that the set of feasible events becomes configuration-dependent and non-locally coupled. A single microscopic update may simultaneously create or remove $\mathcal{O}(r)$ potential events, where $r$ is the hyperedge size, making standard local bookkeeping strategies ineffective.

The algorithms introduced here are designed to resolve this structural bottleneck by reorganizing the event space and its sampling in a way that preserves statistical exactness while restoring near-linear computational scaling. As a result, they enable systematic exploration of discontinuous transitions, hysteresis, and heterogeneous higher-order systems at scales that were previously inaccessible. This strategy drastically reduces computational complexity by replacing the global update of event lists with local sampling rules, allowing the simulation of hypergraphs with millions of nodes and broad distributions of both connectivity and order. We demonstrate the performance of the algorithms using the Susceptible$-$Infected$-$Susceptible (SIS) model with critical mass thresholds, a paradigmatic example of higher-order contagion dynamics~\cite{deArruda2024}. Using the Susceptible$-$Infected$-$Recovered (SIR) epidemic model, we illustrate how the method can easily be adapted to other spreading processes. Beyond epidemic modeling, the proposed methods offer a general framework for scalable, continuous-time simulations of spreading and activation processes in complex systems with many-body interactions.

\section*{Results}
\subsection*{Definition and concepts}

A convenient mathematical representation of higher-order interactions is the hypergraph, $\mathcal{H}=\{\mathcal{N},\mathcal{E}\}$, where $\mathcal{N}=\{i_1,i_2,\ldots,i_N\}$ denotes the set of nodes and $\mathcal{E}=\{h_1,h_2,\ldots,h_H\}$ the set of hyperedges. Each hyperedge $h$ connects $m_h\!+\!1$ nodes and represents an $m_h$-order interaction: first-order hyperedges correspond to pairwise links, second-order to triads, and so on. A special case is that of simplicial complexes~\cite{Torres}, where every $m$-order interaction implies the presence of all lower-order interactions $1,2,\ldots,m-1$, enforcing a hierarchical inclusion structure that general hypergraphs do not require.

Concepts originally defined for pairwise networks can be generalized to higher-order structures~\cite{Battiston2021,Bick}. The generalized degree or $m$-degree of node $i$, denoted $k_i(m)$, is the number of $m$-order hyperedges that include it. The node’s hyperdegree vector is then $\mathbf{k}_i = \{k_i(1), k_i(2), k_i(3), \ldots\}$, and the total number of interactions involving node $i$ is $K_i=\sum_m k_i(m)$. Statistical descriptors such as degree distributions  can be computed for each order, providing a complete structural characterization of higher-order connectivity~\cite{Battiston2021}.

\subsection*{Hyper-SIS dynamics}
\label{sec:Processes}

Contagion processes encompass a wide range of spreading phenomena, from disease transmission to information and opinion dynamics~\cite{Barrat,Castellano,Pastor-Satorras2015,DeArruda2018PR}. Here we consider the susceptible–infected–susceptible (SIS) dynamics on hypergraphs (Hyper-SIS), where each node $i$ can be in one of two states: susceptible ($\sigma_i=0$) or infected ($\sigma_i=1$). Each hyperedge $h$ can be active ($\zeta_h=1$) if its number of infected nodes exceeds a critical mass threshold, or inactive ($\zeta_h=0$) otherwise. Infection events occur within hyperedges at rates $\beta(m)$, which are functions of the orders of the hyperedges to which the node belongs, and infected nodes recover spontaneously at a rate $\alpha$, returning to the susceptible state~\cite{Cencetti,deArruda2024}.

\begin{figure}[hbt]
	\begin{center}
		\includegraphics[width=0.90\linewidth]{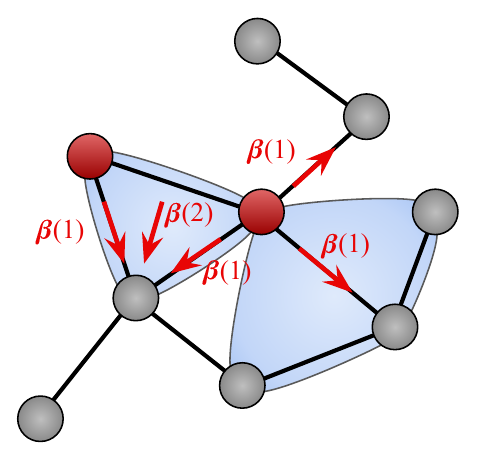}
	\end{center}
	\caption{\justifying \textbf{Schematic representation of the Hyper-SIS model.} Infected nodes (red) transmit the disease to susceptible nodes (gray) via pairwise edges or higher-order interactions, the latter depicted as light-blue shaded regions. In this schematic model, hyperedges are active only if their number of infected nodes exceeds a critical mass threshold $\theta(m)=m$. Pairwise infections occur at a rate $\beta(1)$, second-order infections at a rate $\beta(2)$, and so on. Note that no infection with rate $\beta(3)$ is depicted since the hyperedge is not active.
    }
	\label{fig:Contagion}
\end{figure}

We focus on the Hyper-SIS model with a critical mass threshold~\cite{Barrat2022}, where contagion in a hyperedge of order $m$ is possible only if at least $\theta{(m)}$ of its $m+1$ members are infected. Each susceptible node in an active hyperedge is then infected independently at rate $\beta(m)$.  We denote the vectors of rates and of thresholds by $\boldsymbol{\beta} = \{\beta(1), \beta(2),\ldots,\beta(m)\}$ and $\boldsymbol{\theta}= \{\theta(1), \theta(2),\ldots,  \theta(m)\}$, respectively. Figure~\ref{fig:Contagion} schematically illustrates the process. More general activation mechanisms of hyperedges are discussed elsewhere~\cite{deArruda2024}.

We use a simple rule for infection that enhances the higher-order interaction effect assuming an infection rate that increases linearly with the hyperedge order:
\begin{equation}
	\beta{(m)} = \beta[1+b(m-1)],
\end{equation}
where the pairwise infection rate is $\beta{(1)} = \beta$, and a factor $b$ controls how the interaction strength increases with the hyperedge order. Other relations have been investigated, such as a logarithmic dependence~\cite{Barrat2022} and, more frequently, a restriction to pairwise and triadic interactions where $\beta{(m)}=0$ for $m>2$~\cite{Landry, Iacopini}. 
We also assume a simple linear dependence for the critical mass threshold:
\begin{equation}
	\theta(m) = 1 + (m-1)\theta_0.
\end{equation}
For $\theta_0 = 0$, a single infected agent is sufficient to activate the hyperedge, while for $\theta_0 = 1$, the activation occurs when all but one agent are infected.

This model naturally reproduces the rich phenomenology of higher-order contagion, including continuous and discontinuous transitions, bistability, and hysteresis, depending on the order composition and threshold parameters. In the following sections, we use the Hyper-SIS dynamics as a benchmark for assessing the accuracy and efficiency of the proposed Gillespie algorithms.

\subsection*{Standard Gillespie algorithm for Hyper-SIS}
\label{sub:GA}

In a Markovian contagion dynamics, healing and infection events are modeled as a sequence of Poisson processes associated with changes in the state of nodes or edges~\cite{Daley1999}.
Standard and optimized Gillespie algorithm for simulating Markovian SIS dynamics on simple graphs, originally introduced elsewhere~\cite{Cota2017}, is presented in the "Methods" section. This algorithm serves as the foundation for the extensions developed here to handle higher-order interactions.

The Hyper-SIS model retains the same implementation of the healing dynamics of the pairwise model, where a list of all infected nodes, $\Lambda^{(\text{I})}$, is continuously updated, yielding the total healing rate given by (see the "Methods" section)
\begin{equation}
	F = \sum_{i=1}^N \alpha \sigma_i =  \alpha N_\text{I},
	\label{eq:HealRate}
\end{equation}

\noindent where $N_\text{I}$ is number of infected agents.

In contrast, updating the infection processes becomes substantially more demanding as higher-order interactions are introduced, since the number of possible infection events grows rapidly with both node degree and hyperedge order. In what follows, we present alternative strategies to implement Hyper-SIS dynamics on hypergraphs with arbitrary degree and order distributions, beginning with lower algorithmic complexity and efficiency, and progressing toward more sophisticated algorithms that achieve higher computational performance. The main symbols used to describe the Hyper-SIS dynamics are summarized in Table~\ref{tab:symbols}.

\begin{table}[htb]
	\centering
	\caption{\justifying List of the main symbols used in the algorithms for simulations of Hyper-SIS. Acronym: HE (hyperedge).}
	\resizebox{\linewidth}{!}{%
		\small
		\begin{tabular}{cc}
			\hline\hline
			State of node $i$ & $\sigma_i$ \\
			State of HE $h$ & $\zeta_h$ \\
			Quiescence state of node $i$ & $\eta_i$ \\
			List of infected nodes & $\Lambda^{(\text{I})}$ \\
			List of susceptible nodes in active HEs & $\Lambda^{(\text{S})}$ \\
			List of potentially active HEs & $\Lambda^{(\text{H})}$ \\
			List of potentially quiescent nodes & $\Lambda^{(\text{Q})}$ \\
			\makecell{Maximum number of susceptible nodes within\\ an active HE} & $\omega_h$ \\
			Number of susceptible nodes in HE $h$ & $n^{(\text{S})}_h$ \\
			\makecell{Number of active HE of order $m$ that \\contain node $i$}  & $n_i{(m)}$ \\
			Number of quiescent nodes & $N_Q$ \\
			Number of nodes & $N$ \\
			Number of HEs & $H$\\
			Interaction distribution & $P_K$\\
			Order distribution & $f_m$ \\
			Power-law interaction distribution exponent & $\gamma_k$ \\
			Power-law order distribution exponent & $\gamma_m$ \\
			Hyperdegree of a node $i$ & $\mathbf{k}_i$\\
			$m$-degree of a node $i$ & $k_i(m)$\\
			Order of HE $h$ & $m_h$\\
			Total healing rate & $F$\\
			Total infection rate & $G$\\
			Infection rate in a HE of order $m$ & $\beta(m)$\\
			Critical mass threshold in a HE of order $m$ & $\theta(m)$\\
			\hline
			\hline
		\end{tabular}
	}
	\label{tab:symbols}
\end{table}

Since contagion can occur in any active hyperedge $h$ of order $m_h$, the standard GA for Hyper-SIS requires a list  $\Lambda^\text{(S)}$ of 2-tuples $(i,h)$ with all susceptible nodes $i$ belonging to active hyperedges $h$. The total infection rate is given by
\begin{equation}
    G = \sum_h \beta(m_h) \left[ \sum_{i \in h} (1-\sigma_i) \right] \zeta_h.
    \label{eq:HOInfectRate}
\end{equation}
The standard GA algorithm proceeds as follows. At each time step, the lists $\Lambda^{(\text{I})}$ and $\Lambda^{(\text{S})}$ are constructed, and the total rate of events is computed as $R = F + G$. An infection or a healing event is then selected with probabilities $f = F/R$ and $1 - f = G/R$, respectively. If a healing event is chosen, an infected node is randomly selected from $\Lambda^{(\text{I})}$ and healed. If an infection event is selected, a susceptible node belonging to an active hyperedge $h$ is drawn from $\Lambda^{(\text{S})}$ with probability proportional to $\beta(m_h)$ and infected. After each event, time is advanced by $\tau = -\ln u / R$, and the states of the affected nodes ($\sigma_i$) and hyperedges ($\varsigma_h$) are updated accordingly.

Rebuilding the infection and recovery lists differs substantially in computational cost. Updating $\Lambda^{(I)}$ after a healing event is algorithmically simple and inexpensive, as it does not depend on network connectivity: the healed node can be removed from $\Lambda^{(\text{I})}$ by replacing its entry with the last element of the list. In contrast, updating $\Lambda^{(\text{S})}$ is considerably more demanding. Following either an infection or healing,  $\Lambda^{(\text{S})}$ must be scanned sequentially to remove all deactivated entries. Concomitantly, all hyperedges containing a newly infected node are scanned appending any that became active to $\Lambda^{(\text{S})}$. A simpler but computationally prohibitive alternative is to reset both $\Lambda^{(\text{I})}$ and $\Lambda^{(\text{S})}$ by visiting all nodes and hyperedges after every event. To distinguish these implementations, we refer to the algorithm with full list rebuilding as GA and to the local-update version as GA$^+$. Although the former is inefficient, it is often used because of its algorithmic simplicity and serves as a reliable benchmark for validating optimized implementations.

\subsection*{Hyperedge-based optimized algorithm (HB-OGA)}
\label{sub:HBOGA}

\begin{figure*}
    \begin{center}
		\includegraphics[width=0.65\textwidth]{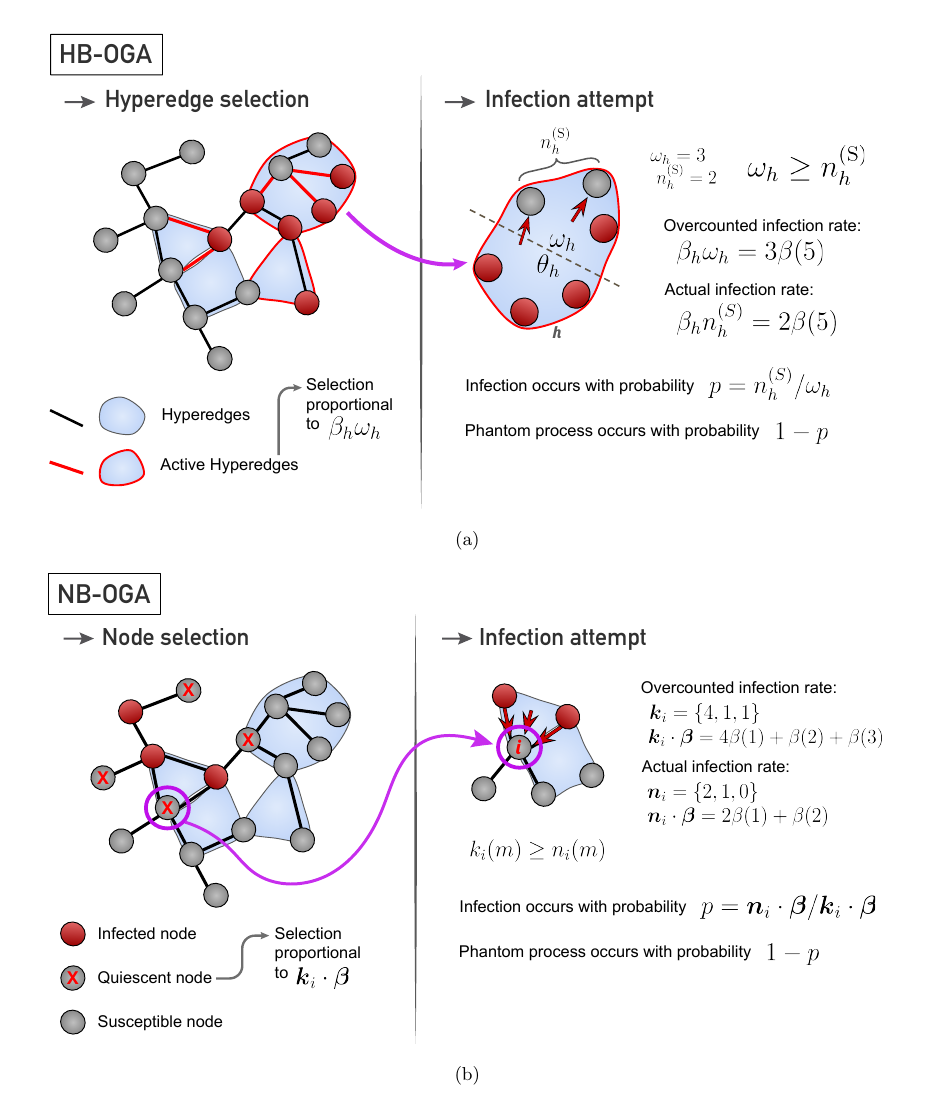}
	\end{center}
	\caption{\justifying \textbf{Schematic representation of optimized algorithms for hyper-SIS dynamics.} (a) Hyperedge-based (HB) algorithm. A hyperedge (shaded regions) is active (red border) when the number of active nodes (red) satisfies the critical-mass threshold condition. Active hyperedges are selected with probability proportional to their maximum infection rate ($\beta_h \omega_h$), while the ratio between actual and overcounted infection rates determines the probability of a phantom process. (b) Node-based (NB) algorithm. Quiescent nodes (crossed circles), i.e., susceptible nodes belonging to active hyperedges, are selected with probability proportional to their maximum allowed infection rate ($\mathbf{k}_i \cdot \boldsymbol{\beta}$). The actual infection rate is then recovered via the corresponding phantom process probability. See Table~\ref{tab:symbols} for a description of the symbols used.}	
	\label{fig:scheme}
\end{figure*}

The high computational cost of creating and managing infection lists can be reduced by introducing a list of potentially active hyperedges, $\Lambda^{(\text{H})}$, whose update has low computational and algorithmic complexity. Let $n_h^{(\text{S})}$ denote the number of susceptible nodes in hyperedge $h$. Instead of the exact potential infection of $n_h^{(\text{S})}$ susceptible nodes, assume that each active hyperedge $h$ attempts to infect $\omega_h = m_h + 1 - \theta(m_h)$ nodes at rate $\beta(m_h)$, where $\omega_h$ represents the maximum number of susceptible nodes allowed within an active hyperedge of order $m_h$, given that at least $\theta(m_h)$ nodes must be infected for activation. The total infection attempt rate is therefore
\begin{equation}
	G = \sum_{h \in \Lambda^{(\text{H})}} \beta(m_h)\, \omega_h,
	\label{eq:HOInfectRate2}
\end{equation}
where the sum may include inactive hyperedges ($\zeta_h = 0$) still present in the list $\Lambda^{(\text{H})}$, leading to an overestimation of the true infection rate in Eq.~\eqref{eq:HOInfectRate}.

The algorithm proceeds analogously to the optimized Gillespie algorithm (OGA) originally used in pairwise epidemic processes~\cite{Cota2017} (see the section "Methods" for details). A healing or infection attempt event is selected with probabilities $f = F/R$ and $1 - f = G/R$, respectively, using Eqs.~\eqref{eq:HealRate} and~\eqref{eq:HOInfectRate2}. For a healing event, an infected node $i$ is randomly chosen from $\Lambda^{(\text{I})}$ and its state is set to susceptible. This update modifies $\sigma_i$, the list of infected nodes $\Lambda^{(\text{I})}$, and the number of susceptible members $n_h^{(\text{S})}$ for all hyperedges containing node $i$. To avoid the costly task of scanning the entire list, neither $\Lambda^{(\text{H})}$, the total attempt rate $G$, nor the hyperedge states $\zeta_h$ are updated at this stage.

For infection events, a hyperedge $h$ is randomly selected from $\Lambda^{(\text{H})}$ with probability proportional to $\beta_h \omega_h$. If $h$ is inactive, the total attempt rate $G$ is adjusted, $h$ is removed from the list, and a phantom process occurs effectively cleaning $\Lambda^{(\text{H})}$ without a full sweep. If $h$ is active, one of its susceptible nodes becomes infected with probability $n_h^{(\text{S})}/\omega_h$; otherwise, with complementary probability $1 - n_h^{(\text{S})}/\omega_h$, a phantom process occurs. In both cases, time advances by $\tau = -\ln u / R$. These rejection events exactly compensate for the overcounting of potential infections, ensuring that the algorithm remains statistically exact. The schematic representation of the HB-OGA algorithm is summarized in Fig.~\ref{fig:scheme}(a).

\subsection*{Node-based optimized algorithm (NB-OGA)}
\label{sub:NBOGA}

An alternative optimization strategy tracks the infection attempts at the level of individual nodes rather than hyperedges. In this node-based approach, the algorithm monitors susceptible nodes that belong to at least one active hyperedge  -- hereafter called quiescent nodes. Let $\mathbf{n}_i = \{n_i{(1)},n_i{(2)},\ldots,n_i{(m)}\}$ denote the vector giving the number of active hyperedges of each order that include node $i$. The corresponding infection rate for node $i$ is $\boldsymbol{\beta}\cdot \textbf{n}_i$. A node is quiescent if $||\textbf{n}_i||>0$, represented by the variable $\eta_i = 1$, and non-quiescent otherwise ($\eta_i = 0$).

The algorithm assumes that each quiescent node $i$ experiences infection attempts at rate $\boldsymbol{\beta}\cdot \textbf{k}_i \ge \boldsymbol{\beta}\cdot \textbf{n}_i$. Thus, all hyperedges containing node $i$ may attempt to infect it, regardless of whether they are currently active. The total infection attempt rate is
\begin{equation}
	G = \sum_i \left(\boldsymbol{\beta}\cdot \textbf{k}_i\right) \eta_i,
	\label{eq:HOInfectRate3}
\end{equation}
which overestimates the true infection rate. The excess is corrected statistically by phantom processes, as described below.

Moreover, at each step, a healing or infection event is selected with probabilities $f = F/R$ and $1 - f = G/R$, using Eqs.~\eqref{eq:HealRate} and~\eqref{eq:HOInfectRate3}. Healing events follow the same procedure as before: an infected node is randomly chosen from $\Lambda^{(\text{I})}$, its state $\sigma_i$ is changed to susceptible, and its quiescence state $\eta_i$ is updated, subsequently adding the node to the list of quiescent nodes $\Lambda^{(\text{Q})}$ if necessary, but the quiescence of its neighbors is not updated at this point.  For infection events, the algorithm constructs $\Lambda^{(\text{Q})}$, the list of all $N_\text{Q}$ nodes that may belong to an active hyperedge. A node is selected from $\Lambda^{(\text{Q})}$ with probability proportional to $\boldsymbol{\beta} \cdot \mathbf{k}_i$. If the node is found to be non-quiescent, it is removed from the list, $G$ is updated, and a phantom process occurs. Otherwise, node $i$ is infected with probability $\boldsymbol{\beta} \cdot \mathbf{n}_i/\boldsymbol{\beta} \cdot \mathbf{k}_i$; if the attempt fails, a phantom process takes place. In both cases, the simulation time advances by $\tau = -\ln u / R$. This node-based scheme significantly lowers computational complexity compared with a full reset of infection lists, since only the neighborhood of the selected node is inspected to update $\mathbf{n}_i$. The list $\Lambda^\text{(Q)}$ can also be maintained efficiently: when nodes become quiescent, they are appended to the list, while those that lose all active hyperedges remain temporarily until selected during later infection attempts. The schematic representation of the NB-OGA algorithm is summarized in Fig.~\ref{fig:scheme}(b).

\subsection*{Validation of the optimized algorithms}

\begin{figure*}[thb]
	\begin{center}
		\includegraphics[width=0.65\textwidth]{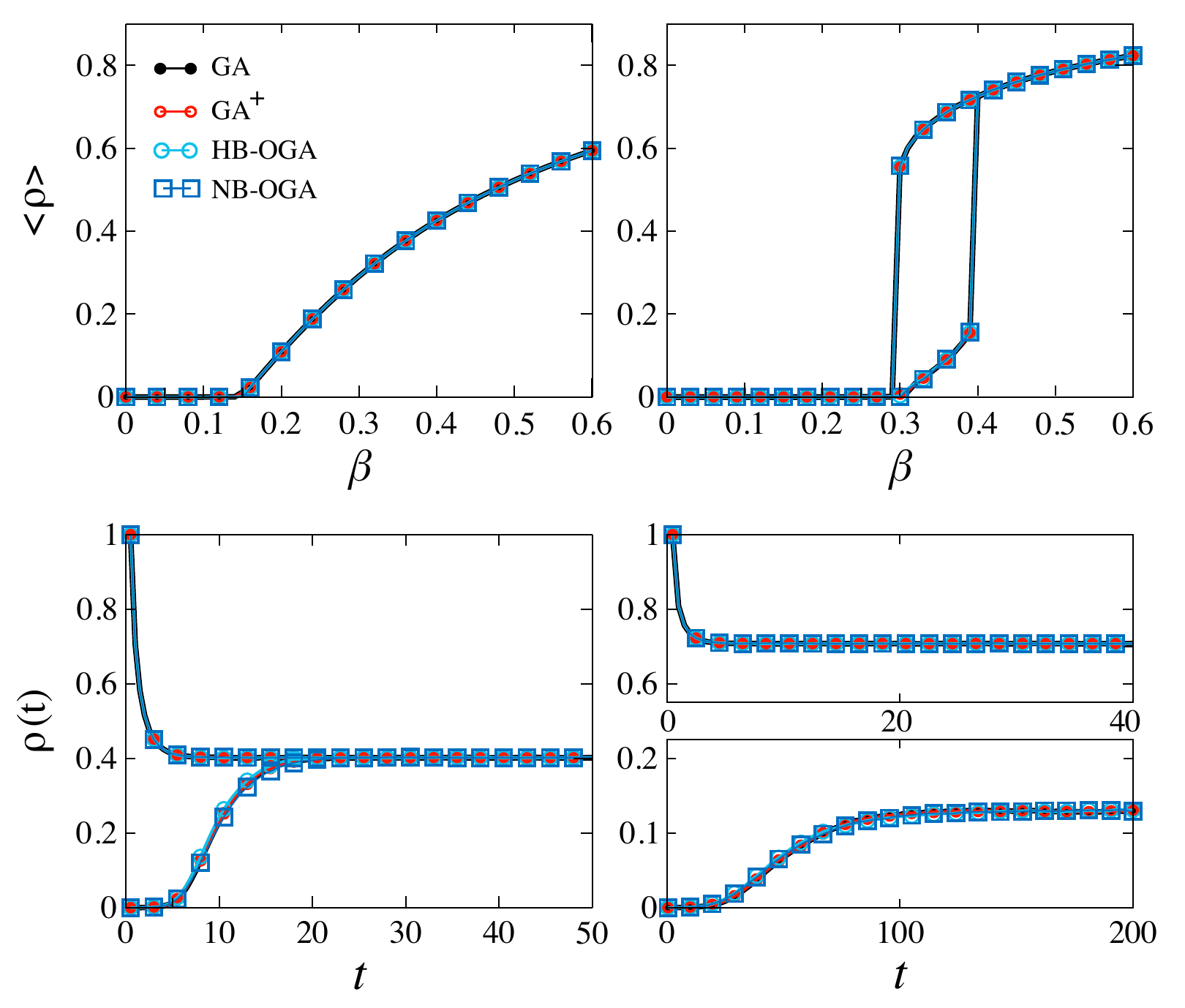}
	\end{center}
	\caption{\justifying \textbf{Comparison of the Hyper-SIS dynamics run with different algorithms.} The top panels show the stationary prevalence, $\langle\rho\rangle$, as a function of the pairwise infection rate, $\beta(1)=\beta$, while the bottom panels show the time evolution of the prevalence, $\rho(t)$, for different initial conditions. Simulations were run for 200 independent realizations for networks with $N=16,000$ nodes. The interaction distribution follows a power law $P_K \sim K^{-2.7}$ with $K \in [3, K_\text{c}]$, and the order is distributed according to a power law $f_m \sim (m+1)^{-\gamma_m}$ with $m \in [1, m_\text{c}]$. Here, $K_\text{c}$ and $m_\text{c}$ are rigid cutoffs. Parameters $\gamma_m = 6.0$, $b = 0$, and $\theta_0 = 1$ are used in the left hand panel, while $\gamma_m = 3.0$, $b = 0.8$, and $\theta_0 = 0.7$ are used in the right hand panels. Acronyms: GA (standard Gillespie algorithm with full list rebuilding), GA$^+$ (standard Gillespie algorithm with local-update), HB-OGA (hyperedge-based optimized Gillespie algorithm) and NB-OGA (node-based optimized Gillespie algorithm).}
	\label{fig:Comparison}
\end{figure*}

To compare algorithmic performance across sizes and structural regimes, we generate synthetic hypergraphs using the bipartite configuration model (BCM), adapted from Courtney and Bianconi~\cite{Courtney}; see Supplementary Fig. 1. The BCM allows to produce hypergraphs with arbitrary degree and order distributions $P_K$ and $f_m$, respectively. Extensive simulations were performed to assess the efficiency of the proposed algorithms across different network sizes and levels of heterogeneity. Their statistical equivalence is demonstrated in Fig.~\ref{fig:Comparison}, which shows that all methods yield identical results for both the stationary and temporal evolution of the epidemic prevalence. 

Performance was evaluated for three classes of structures: homogeneous hypergraphs, power-law hypergraphs, and hypergraphs containing hyperblobs~\cite{Barrat2022}. For each network type and size, an ensemble of 20 realizations was generated and sampled once. The total simulation time was fixed at $T = 10{,}000~\alpha^{-1}$. When the system reached an absorbing state, a single node was randomly reactivated following the quasistationary reactivation method of Ref.~\cite{Costa2021}.

For uniform interaction distributions, the epidemic transition points are approximately invariant with network size, whereas heterogeneous contact distributions exhibit pronounced finite-size effects. The parameter sets were chosen to explore the relevant dynamical regimes by fixing all parameters except $\beta(1)$, which was adjusted to maintain a constant epidemic prevalence as network size increased.

All CPU times were measured on a workstation equipped with an Intel Core i7-13700F processor (5.2~GHz), 64~GB of DDR4 RAM, and a 250~GB NVMe SSD. The code was implemented in Fortran and compiled with version~2024.2.0 of the LLVM-based Intel Fortran compiler (\texttt{ifx}) for 64-bit Linux, using double precision and standard optimization flags.

\subsection*{Computational performance for homogeneous number of interactions}

A simple example of a hypergraph with a homogeneous distribution of interactions is obtained by assigning the same number of interactions, $K = 8$, to every node, while controlling the fraction of hyperedges of each order $m$ through the order distribution $f_m$. Specifically, we set $f_1 = 1/2$ for pairwise connections, $f_m = 1/8$ for hyperedges of order $m = 2, 3, 4,$ and $5$, and $f_m = 0$ otherwise. Although each node has an identical total number of interactions, its generalized degree is heterogeneous because it depends on the orders of the attached hyperedges. The average generalized degrees are $\langle k_i(1)\rangle = 4$, $\langle k_i(m)\rangle = 1$ for $m = 2, 3, 4, 5$, and $k_i(m) = 0$ otherwise. This parameter choice ensures that higher-order effects are non-negligible while keeping interactions evenly distributed across nodes.

\begin{figure*}[htbp]
    \centering
    \includegraphics[width=0.7\textwidth]{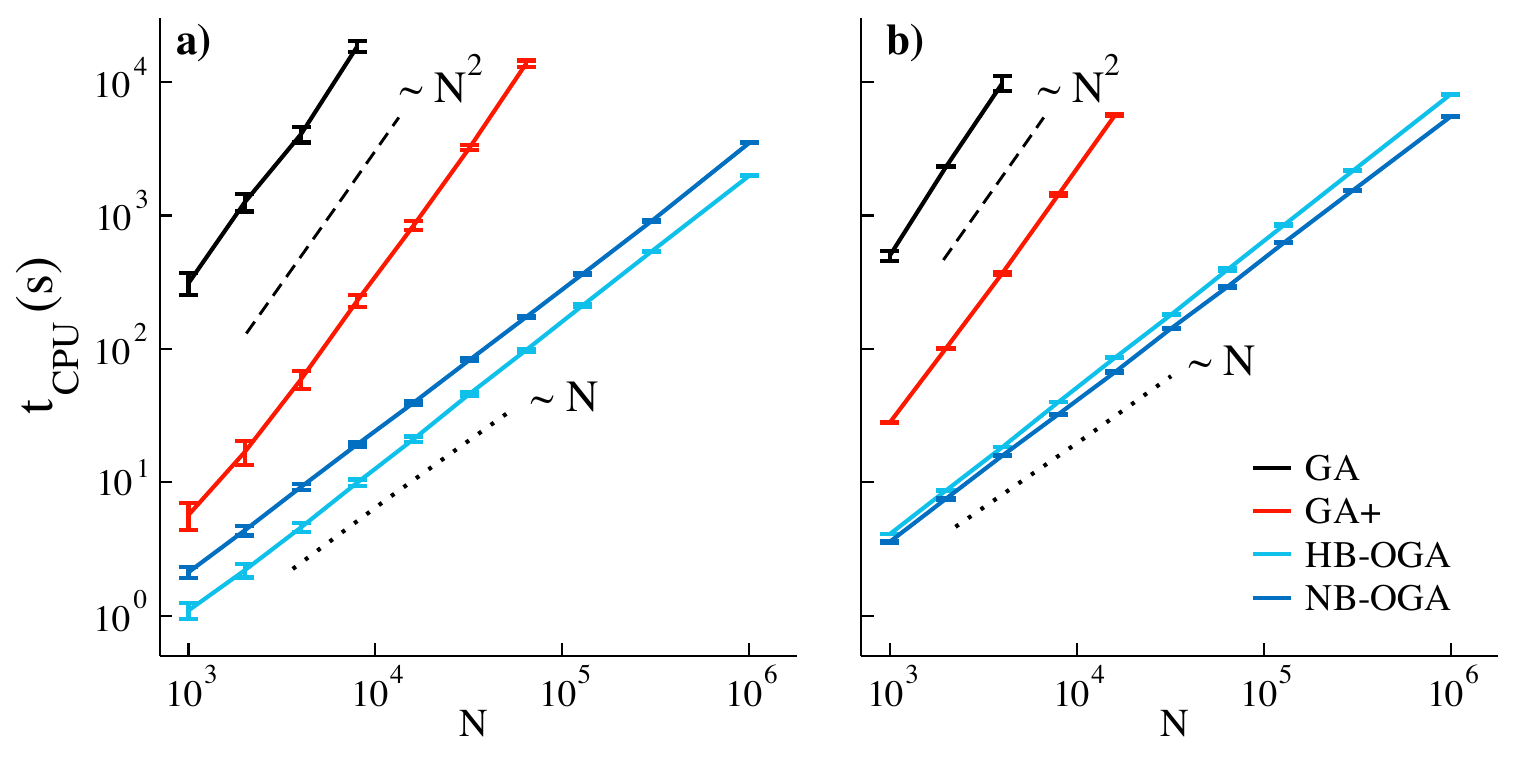}

    \caption{\justifying \textbf{Algorithms' performance on homogeneous contact distributions.} CPU times for the simulation of Hyper-SIS contagion dynamics on a hypergraph with $\langle k(1) \rangle = 4$ and $\langle k(m) \rangle = 1$ for $m=2$ to $5$. First-order spreading rates $\beta{(1)}$ were selected to maintain a stationary prevalence $\rho \approx 0.3$ with $b = 0$ in (a) and $\rho \approx 0.82$ with $b = 1$ in (b); the former was achieved through an initial condition starting from a single infected node while the latter starting with all nodes infected. Parameters were chosen to produce a continuous phase transition in (a) and a discontinuous phase transition in (b). Critical mass threshold is determined by $\theta_0 = 1$. Error bars are defined as the standard deviation of the CPU time obtained from 20 sampled networks. Acronyms are defined in Figure~\ref{fig:Comparison}.}
    \label{fig:CPU_Hom}
\end{figure*}

For the contagion dynamics, we analyze two representative regimes: one with low higher-order spreading rates, dominated by pairwise interactions, and another with strong higher-order effects. The first regime exhibits a continuous transition from the disease-free to the endemic state, as expected in pairwise contagion models~\cite{Pastor-Satorras2001}, whereas the second regime, dominated by higher-order interactions, displays a discontinuous transition from the disease-free to a highly active epidemic phase~\cite{deArruda2021}. Figure~\ref{fig:CPU_Hom} shows the CPU times (in seconds) as a function of network size, comparing the efficiency of the algorithms in the supercritical regime for both continuous and discontinuous transitions.

As expected, the non-optimized implementation, which fully rebuilds the lists of possible processes after each event in GA, exhibits a high computational complexity that scales as $\sim N^2$. Nevertheless, it serves as a benchmark to verify the correctness of the more advanced and efficient algorithms, as shown in Fig.~\ref{fig:Comparison}. The local-update strategy implemented in GA$^{+}$ is approximately 80 times faster in the continuous regime and about 20 times faster in the discontinuous regime, although its overall computational complexity still scales as $N^2$. In contrast, the introduction of phantom processes in both HB-OGA and NB-OGA drastically reduces computational complexity, yielding CPU times that scale nearly linearly with system size. Despite their similar scaling, the hyperedge-based approach outperforms the node-based method in low-prevalence regimes, where its CPU times are roughly half those of NB-OGA. Conversely, in high-prevalence regimes, NB-OGA becomes slightly more efficient than HB-OGA. As discussed in for power-law distributions, this relative performance may invert again when strong structural heterogeneity is present.

\subsection*{Computational performance for power-law distributed hypergraphs}
\label{subsec:PL}

The nearly linear computational complexity observed for homogeneous distributions remains valid for heterogeneous order and interaction distributions when they have a size-independent upper cutoff. When the distribution's upper cutoffs increase with size, such as in power-laws, the algorithm performance has additional demands and linear computation complexity is no longer guaranteed. To evaluate the algorithms’ performance in networks with both interaction and order heterogeneities, we select power-law distributions, $P_K \sim K^{-\gamma_k}$ and $f_{m} \sim (m+1)^{-\gamma_m}$, respectively. To reduce sample-to-sample fluctuations and avoid outliers in randomly generated networks with heavy-tailed distributions, we impose rigid cutoffs $K_\text{c}$ and $m_\text{c}$, defined by $N P_{K_\text{c}} = 1$ and $H f_{m_\text{c}} = 1$, which yields $K_\text{c} \sim N^{1/\gamma_k}$ and $m_\text{c} \sim H^{1/\gamma_m}$, still scaling with network size. This procedure suppresses large fluctuations in the maximum number of interactions and the highest hyperedge order, which could hinder precise comparisons of computational efficiency for a finite ensemble of networks~\cite{Silva2019,Boguna2004}. Outliers in order are discussed below in the context of hyperblobs and improved sampling  methods to mitigate their effects are considered in the sequence.

\begin{figure}[htbp]
    \begin{center}
		\includegraphics[width=0.43\textwidth]{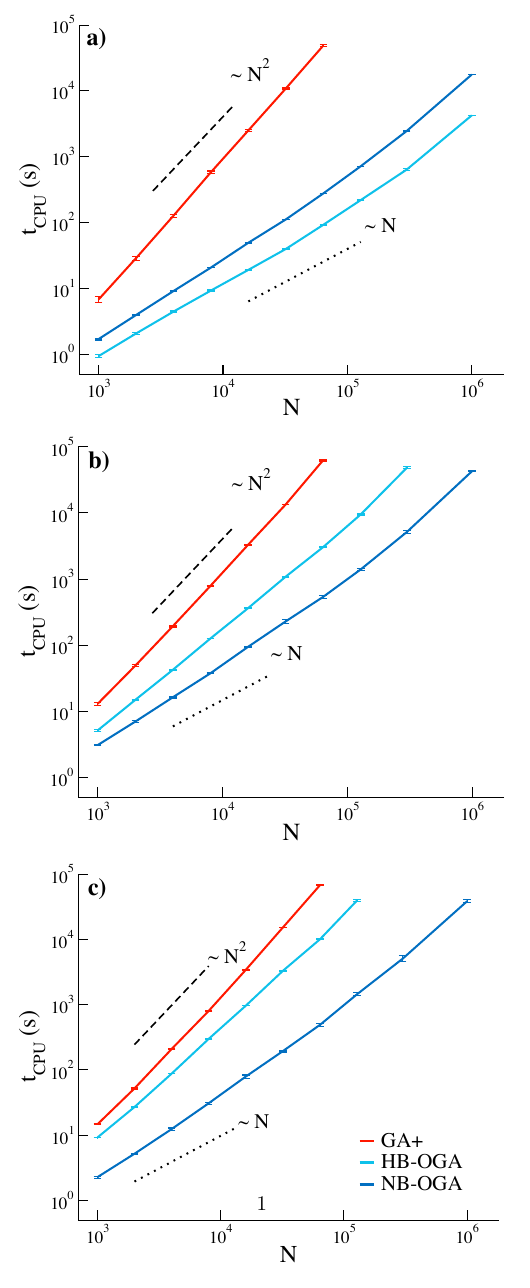}
	\end{center}
    \caption{\justifying \textbf{Algorithms' performance on power-law distributed hypergraphs.} CPU times for simulations of the Hyper-SIS contagion on hypergraphs with power-law interaction and order distributions. First-order spreading rates, $\beta(1)$, were chosen to maintain the same stationary epidemic prevalence on networks of different sizes. Exponents (a) $\gamma_m = 6$ and $\rho \approx 0.3$; (b) $\gamma_m = 3$ and $\rho \approx 0.67$; and (c) $\gamma_m = 2.5$ and $\rho \approx 0.8$ were considered. Heterogeneous interaction distribution was considered for fixed $\gamma_k = 2.7$ and the degree range $K \in [3, K_\text{c}]$, where $K_\text{c}(N)$ is a rigid cutoff. Acronyms are defined in Figure~\ref{fig:Comparison}. }
	\label{fig:CPU_PL}
\end{figure}

We fixed $\gamma_k = 2.7$, minimal degree  $K=3$, rigid cutoff, and examined the effect of order heterogeneity by considering three order distributions: $\gamma_m = 6.0$ with a cutoff at $m_c = 10$, and $\gamma_m = 3.0$ and $2.5$ with rigid cutoffs. Decreasing $\gamma_m$ increases order heterogeneity, ranging from networks dominated by pairwise interactions (approximately 90\% pairwise and 8\% second-order, \ldots) for $\gamma_m = 6.0$, to networks where higher-order interactions are more frequent (around 50\% pairwise, 20\% second-order, 10\% third-order, \ldots) for $\gamma_m = 2.5$.

For $\gamma_m = 6.0$, we set $b = 0$ and $\theta_0 = 1$, yielding dynamics dominated by pairwise interactions and a continuous phase transition. For $\gamma_m = 3.0$, parameters $b = 0.8$ and $\theta_0 = 0.7$ were chosen to produce discontinuous transitions in a regime with moderate higher-order effects. Finally, for $\gamma_m = 2.5$, we used $b = 0.9$ and $\theta_0 = 0.6$ to generate discontinuous transitions with strong higher-order heterogeneity.

Figure~\ref{fig:CPU_PL} shows the CPU times (in seconds) as a function of network size for the different algorithms and the parameter sets described above. For low higher-order heterogeneity ($\gamma_m = 6.0$, Fig.~\ref{fig:CPU_PL}a), the performance trends are similar to those observed for homogeneous interaction distributions in Fig.~\ref{fig:CPU_Hom}: GA$^{+}$ displays high computational complexity ($t_\text{CPU} \sim N^2$), whereas the optimized algorithms achieve near-linear scaling ($t_\text{CPU} \sim N$) for sizes up to $N\sim 10^5$ followed by a crossover to higher computational complexity, still far more efficient than standard GA. Among them, the hyperedge-based approaches are more efficient, running about three times faster than the node-based implementations. For intermediate order heterogeneity ($\gamma_m = 3.0$, Fig.~\ref{fig:CPU_PL}b), HB-OGA markedly outperforms GA$^{+}$ but,  in turn, is less efficient than NB-OGA. The latter exhibits CPU times that scale slightly faster than linearly with network size, corresponding to low computational complexity, while HB-OGA scales approximately as $t_\text{CPU} \sim N^{1.5}$, indicating a higher computational cost. For strong heterogeneity ($\gamma_m = 2.5$, Fig.~\ref{fig:CPU_PL}c), the qualitative behavior is similar to that for $\gamma_m = 3.0$, but both HB-OGA and NB-OGA display increased computational complexity, with CPU times scaling as $t_\text{CPU} \sim N^{1.7}$ and $t_\text{CPU} \sim N^{1.3}$, respectively. The increased computational complexity is due to the increasing rejections as the upper cutoffs in both order and interactions increase; see Supplementary Fig. 2 and the corresponding discussion in the end of this section.

The performances were further evaluated for fixed network parameters and varying epidemic prevalences. We considered a continuous transition, in which the order parameter varies smoothly with the control parameter $\beta(1)$. Simulations were performed with $b = 0.4$ and $\theta_0 = 1.0$ fixed, while $\beta(1)$ was adjusted to produce low and high prevalences, respectively.  We compare the results for a hypergraph with power-law distributions of both node interactions and hyperedge orders; see Supplementary Fig. 3. At high infection densities, the performance of GA$^{+}$ and HB-OGA deteriorates substantially (up to tenfold slower), whereas NB-OGA is only moderately affected (roughly twice slower). Indeed, at low prevalence, NB-OGA is about one order of magnitude slower than HB-OGA, but the two achieve comparable performance at high prevalence. This crossover  arises from the excess of phantom processes in NB-OGA at low prevalence, where the attempt infection rate greatly exceeds the actual rate, i.e., $\boldsymbol{\beta}_i \cdot \mathbf{k}_i \gg \boldsymbol{\beta}_i \cdot \mathbf{n}_i$. Note that for the parameter set used (low $b$ and high $\theta_0$), lower-order activations are favored. In contrast, for the same order heterogeneity and similar prevalence, as shown in Fig.~\ref{fig:CPU_PL}(b) where NB-OGA outperforms HB-OGA, the parameters with high $b$ and lower $\theta_0$ enhance higher-order interactions. These findings indicate that HB-OGA performs best when lower-order interactions dominate the dynamics, while NB-OGA becomes more efficient in regimes driven by strong higher-order effects.

\subsection*{Computational performance for  hypergraphs with a Hyperblob}
\label{subsec:blob}

Outliers are common in complex systems and play a key role in shaping their dynamics. In pairwise networks, for instance, a few nodes with degree $k \gg \langle k \rangle$ can sustain metastable localized activity, where only a vanishing fraction of the system maintains epidemic activity in the thermodynamic limit~\cite{Boguna2013,Silva2021}. In higher-order networks, outliers may arise either as nodes with exceptionally large numbers of interactions or as hyperedges of very high order, $m \gg \langle m \rangle$. To evaluate algorithmic performance in this latter scenario, we added a hyperedge of order $m = N - 1$, referred to as a hyperblob~\cite{Barrat2022}, to networks with power-law degree and order distributions. The hyperblob alone can induce bistability: once activated, it simultaneously transmits infection to all susceptible nodes at rate $\beta(N - 1)$. In the Hyper-SIS model with a critical mass threshold, these hyperblob-driven effects emerge only when the density of infected nodes exceeds the activation condition $\rho N > \theta(N - 1)$.

\begin{figure*}[htbp]
    \centering
    \begin{subfigure}[b]{\linewidth}
        \centering
        \includegraphics[width=0.7\textwidth]{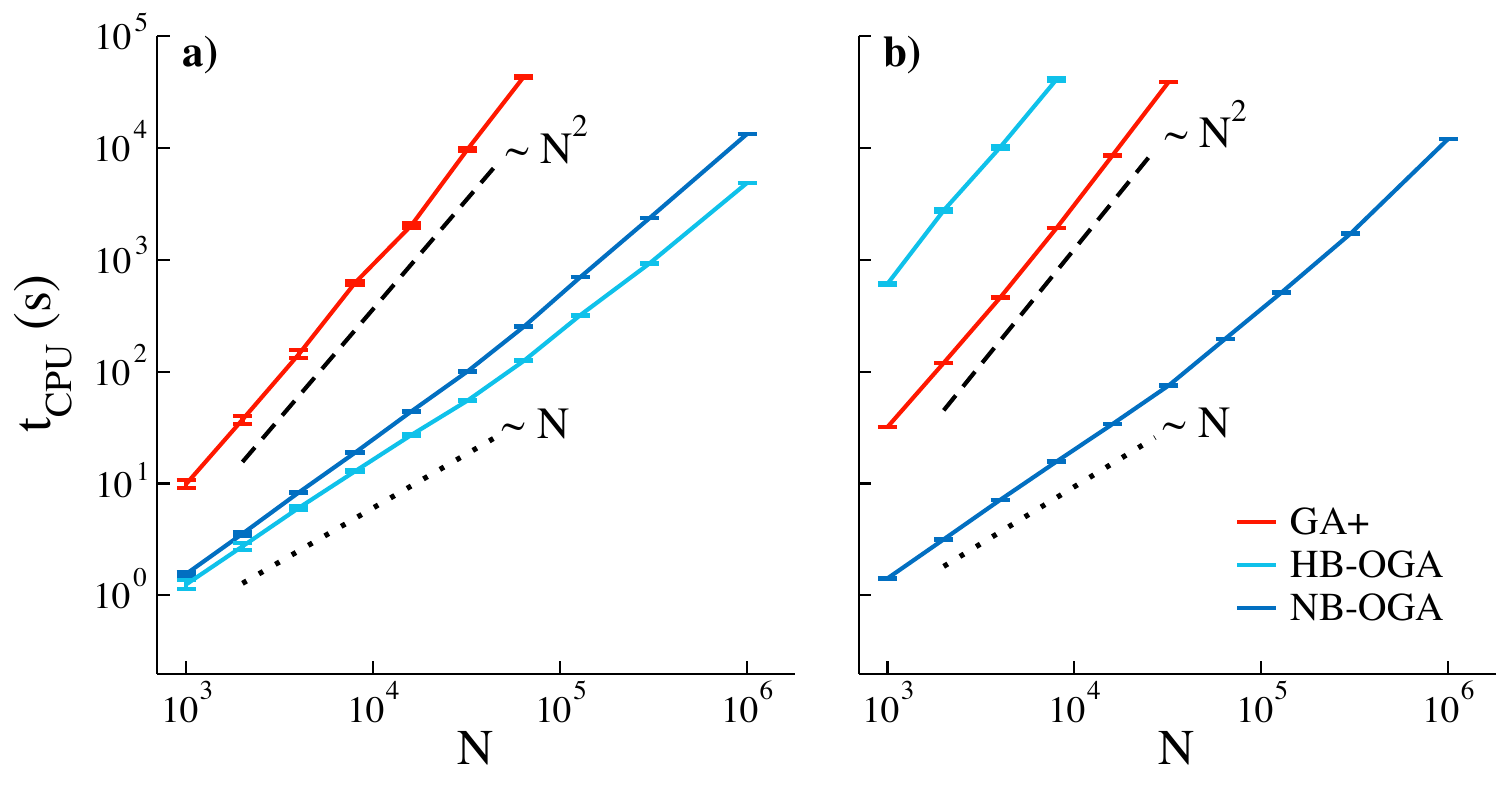}
    \end{subfigure}
  \caption{\justifying \textbf{Algorithms' performance with the presence of a hyperblob.} CPU times for the simulation of Hyper-SIS contagion dynamics on a PL higher-order network of $\gamma_m = 6.0$ plus a hyperblob. The dynamics is bistable and two regimes were tested: (a) lower prevalence where the hyperblob is inactive with $\rho \approx 0.25$  and (b) higher prevalence when the hyperblob is active with $\rho \approx 0.82$. Critical mass threshold is controlled by $\theta_0 = 0.5$ and $\beta(m)=\beta(1)$ for all orders except the hyperblob for which $\beta(N-1) = 20\beta(1)$. Acronyms are defined in Figure~\ref{fig:Comparison}.}
    \label{fig:CPU_blob}
\end{figure*}

The Hyper-SIS dynamics associated with hyperblobs are bistable~\cite{Barrat2022}. To evaluate algorithmic performance, we considered two initial conditions: one where the hyperblob remains inactive ($\rho \approx 0.25$) and another where it is active from the start ($\rho \approx 0.82$). To explore these regimes, a hyperedge of order $N - 1$ was added to networks with $\gamma_k = 2.7$ and $\gamma_m = 6.0$, using the same parameters as in Fig.~\ref{fig:CPU_PL}(a). Figure~\ref{fig:CPU_blob} shows the CPU times (in seconds) obtained for different algorithms applied to the hyperblob case. In the low-prevalence regime, dominated by lower-order interactions, GA$^{+}$ performs inefficiently, exhibiting a computational complexity of $t_\text{CPU} \sim N^2$, whereas HB-OGA and NB-OGA are markedly more efficient, with scaling $t_\text{CPU} \sim N^{1.1}$ and $t_\text{CPU} \sim N^{1.2}$, respectively.

In the high-prevalence regime, dominated by hyperblob activation, HB-OGA becomes substantially slower -- over an order of magnitude less efficient than GA$^{+}$, while NB-OGA retains its computational scaling with a smaller prefactor. The strong slowdown of HB-OGA at high prevalence arises from the excess of rejected attempts when sampling hyperedges in proportion to $\beta_h \omega_h$ once the hyperblob is active. In contrast, NB-OGA improves in this regime because phantom processes become less frequent: the number of active hyperedges $\mathbf{n}_i$ connected to a node $i$ approaches its hyperdegree $\mathbf{k}_i$, bringing the ratio $\boldsymbol{\beta} \cdot \mathbf{n}_i / \boldsymbol{\beta} \cdot \mathbf{k}_i$ closer to unity.

\subsection*{Improved sampling methods}

While the focus of this work is on employing phantom processes to accelerate simulations, the basic rejection method used to sample nodes or hyperedges with broadly distributed weights $\{w_i\}$ can lead to an excessive number of rejections when $w_i / w_\text{max} \ll 1$ for most elements, where $w_\text{max} = \max_i \{w_i\}$. This inefficiency can drastically degrade performance. Sampling efficiency can be improved through binary tree (BT) structures~\cite{DAmbrosio2022}, alias sampling techniques~\cite{DAmbrosio2022}, or -- favoring simplicity -- the improved optimized Gillespie algorithm (IOGA) introduced in Ref.~\cite{Cota2017}. The key idea behind IOGA is to reduce rejections by partitioning quiescent nodes or potentially active hyperedges into two (or more) groups according to their weights: $\Lambda_\text{low} = \{i \mid w_i \le w^*\}$ and $\Lambda_\text{high} = \{i \mid w_i > w^*\}$. The total infection rate is then divided between the two groups,
\begin{equation}
	G_\text{low} = \sum_{i \in \Lambda_\text{low}} w_i,
\end{equation}
and
\begin{equation}
	G_\text{high} = \sum_{i \in \Lambda_\text{high}} w_i.
\end{equation}
When an infection attempt occurs, an element is drawn from $\Lambda_{\text{low}}$ with probability $G_\text{low}/(G_\text{low} + G_\text{high})$, proportionally to its weight; otherwise, an element from $\Lambda_{\text{high}}$ is selected. The rejection step then uses $w_i / w^*$ or $w_i / w_\text{max}$ depending on whether the sample came from the low- or high-weight group, respectively. The threshold $w^*$ should be chosen to minimize the number of rejections. Since weight distributions in epidemic processes typically decay, setting $w^* \gtrsim \langle w \rangle$ is generally sufficient. In our implementation, we adopt $w^* = \langle w \rangle + \sigma_w$, where $\sigma_w = \sqrt{\langle w^2 \rangle - \langle w \rangle^2}$ denotes the standard deviation of the weights.

Binary trees are data structures commonly used to enable efficient weighted sampling in stochastic simulations. In this approach, individual events are stored as leaves, while internal nodes contain the cumulative weights of their descendants, forming a hierarchical representation of partial sums. Sampling is performed by drawing a uniform random number and traversing the tree according to these cumulative weights until a leaf (event) is selected. This procedure, as well as updates to event weights, can be carried out in $\mathcal{O}(\ln N)$ time.

Table~\ref{tab:CPU_table} summarizes the CPU times for Hyper-SIS simulations incorporating both order and contact heterogeneities, confronting the computational performance of standard OGA with IOGA and sampling using binary trees, using the same parameters as in Fig.~\ref{fig:CPU_PL}(b). While IOGA and BT present similar performance with the former being slightly better, their computational advantage become increasingly pronounced as network size and heterogeneity grow. Among the optimized methods, HB-OGA gains more from the IOGA correction because the acceptance probability for hyperedge infection attempts is inversely proportional to $\omega_h = m_h + 1 - \theta(m_h)$, whose distribution is broader than that of the infection attempts for quiescent nodes in NB-OGA. Nevertheless, NB-OGA remains significantly more efficient than HB-OGA in regimes characterized by strong higher-order heterogeneity.

\begin{table*}[hbt!]
	\centering
	
	\caption{\label{tab:CPU_table}\justifying \textbf{Performances of improved rejection methods.} CPU times in minutes to run a total fixed time of $10^4\alpha^{-1}$ in quasistationary simulations of the Hyper-SIS model on a PL higher-order network with degree and order exponents set to $\gamma_k=3$ and $\gamma_m = 3$. Parameters were set to $b=0.6$ and $\theta_0 = 0.8$, with the first-order spreading rate $\beta(1)$ set to maintain a fixed stationary infection density $\rho$ indicated in the table. The rejection methods compared were: the standard rejection OGA, the improved rejection with IOGA and a binary trees (BT).}
	\begin{ruledtabular}
		\begin{tabular}{cccccccc}
			& & \multicolumn{3}{c}{Hyperedge-Based} & \multicolumn{3}{c}{Node-Based} \\
			\cmidrule(lr){3-5} \cmidrule(lr){6-8}
			& $N$ & $10^5$ & $10^6$ & $10^7$ & $10^5$ & $10^6$ & $10^7$ \\
			\midrule
			  & OGA & 7.8 & 252.6 & -- & 8.2 & 284.6 & -- \\
			$\rho = 0.25$ & IOGA & 1.9 & 47.0 & 1121.1 & 6.2 & 176.2 & 4328.6 \\
			& BT & 2.8 & 64.1 & 1665.8 & 8.8 & 181.2 & 4988.0 \\
			\midrule
			& OGA & 40.1 & 2148.5 & -- & 7.3 & 251.7 & -- \\
			  $\rho = 0.8$ & IOGA & 8.4 & 248.2 & 15809.1 & 4.5 & 101.6 & 2971.7 \\
			& BT & 10.5 & 291.0 & 11252.2 & 6.2 & 168.7 & 3678.1 \\
		\end{tabular}
	\end{ruledtabular}
    \label{tab:rejections}
\end{table*}

\subsection*{Algorithms' performance on real networks}

\begin{figure}[htbp]
	\centering
	\begin{subfigure}[b]{\linewidth}
		\centering
		\includegraphics[width=\linewidth]{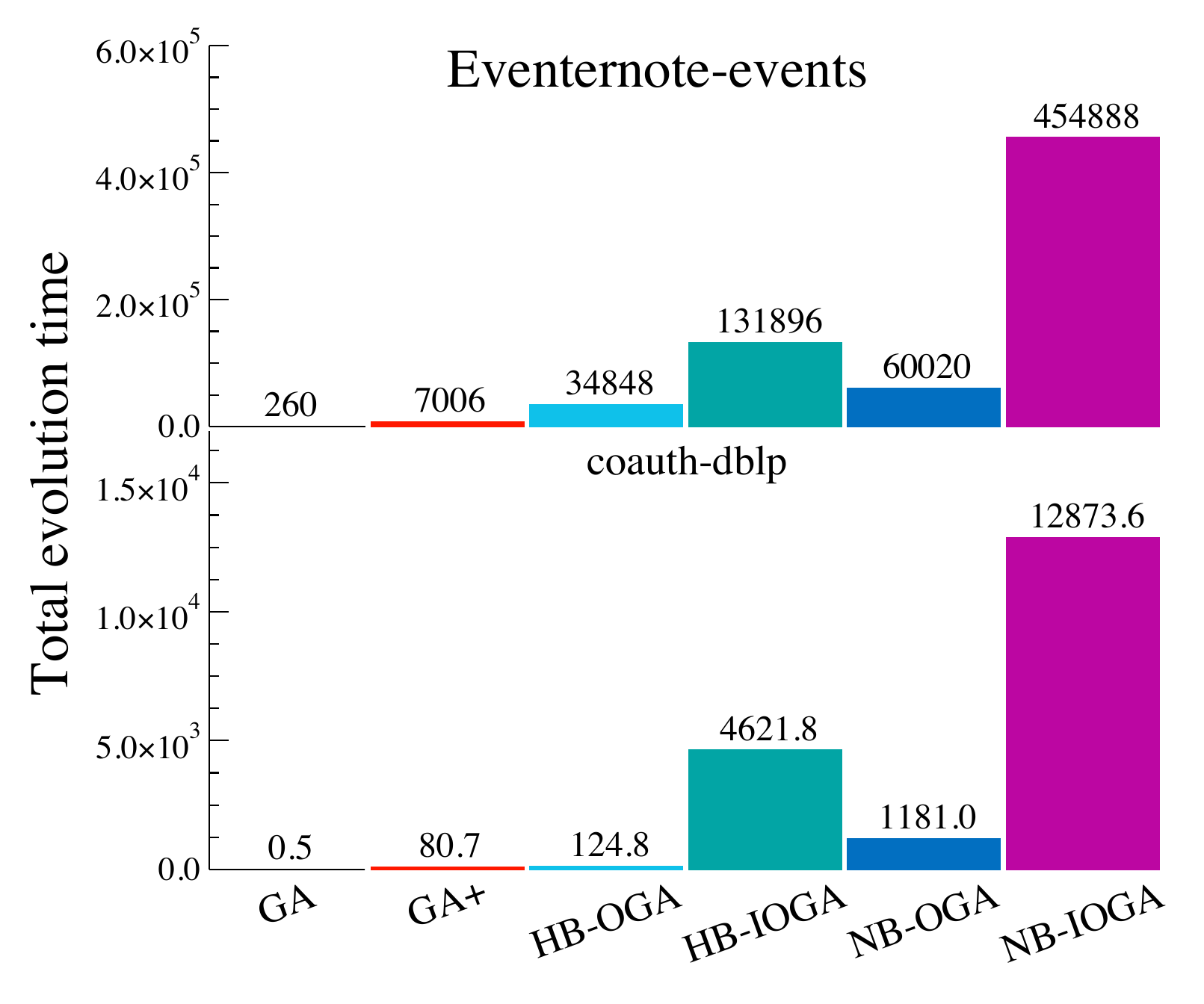}
	\end{subfigure}
	
	\caption{\justifying \textbf{Algorithms' performance on real higher-order networks.} Total evolution time reached after 12 hours of real CPU time of Hyper-SIS on real hypergraphs eventernote-events (69{,}885 nodes and 131{,}647 hyperedges) and coauth-dblp (1{,}659{,}954 nodes and 2{,}093{,}835 hyperedges). For each network, 8 runs were taken with parameters $b = 0.6$, $\theta_0= 0.8$. First-order spreading rates $\beta{(1)} = 0.1$ and $\beta{(1)} = 0.8$ were used for the former and the latter, respectively, in order to obtain a stationary density of infected nodes $\rho \approx 0.4$ in both cases. Acronyms: HB-IOGA (hyperedge-based improved optimized Gillespie algorithms); NB-IOGA (node-based improved optimized Gillespie algorithms); other acronyms are defined in Figure~\ref{fig:Comparison}.}
	\label{fig:TCPU}
\end{figure}

Algorithm performances on real networks are shown in Fig.~\ref{fig:TCPU}. The first dataset, \texttt{eventernote-events}~\cite{Bell2024}, contains 69{,}885 nodes and 131{,}647 hyperedges, where nodes represent individuals and hyperedges correspond to events attended by them. The interaction order ranges from 1 to 619, with an average $\langle m \rangle = 9.05$ and standard deviation $\sigma_m = 13.06$. The second network, \texttt{coauth-dblp}~\cite{Benson2018}, comprises 1{,}659{,}954 nodes and 2{,}093{,}835 hyperedges, with nodes representing authors and hyperedges representing their publications on DBLP. In this case, the interaction order spans from 1 to 219, with $\langle m \rangle = 3.46$ and $\sigma_m = 1.76$. Both datasets were obtained from the XGI Python library~\cite{Landry2023}. The measured CPU times reveal that the standard GA and GA$^{+}$ are much less efficient than the optimized algorithms, with NB-OGA outperforming HB-OGA due to the substantial heterogeneity presented in the real networks. Furthermore, the IOGA further improves efficiency for both approaches, particularly in the larger \texttt{coauth-dblp} network.

\subsection*{Optimized algorithms for other dynamical processes}

\begin{figure*}
	\centering
	\includegraphics[width=0.7\linewidth]{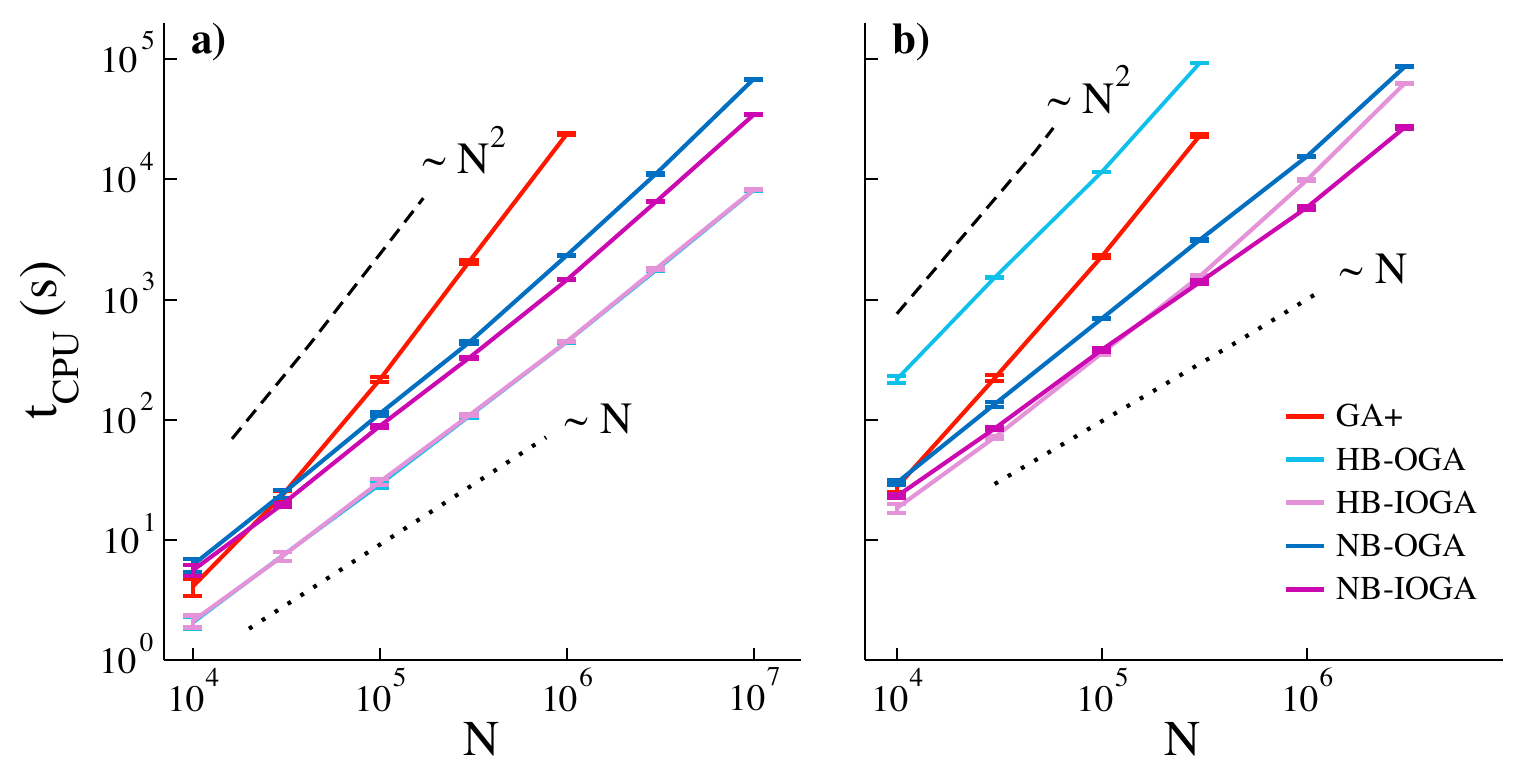}
    \caption{\justifying \textbf{Algorithm performance for SIR spreading on power-law hypergraphs.} CPU times (in seconds) for simulations starting from a single infected node randomly selected and running until activity ceases (i.e., the absorbing state is reached). The higher-order network has power-law degree and order distributions with exponents $\gamma_k = 3$ and $\gamma_m = 3$. Spreading parameters were set to (a) $b=0$, $\theta_0=1.0$, yielding a final recovered fraction $\rho_\text{R} = 0.1$, and (b) $b=0$, $\theta_0=1.0$, yielding $\rho_\text{R} = 0.3$. A total of 10\,000 runs were performed per network. Acronyms are defined in Figure~\ref{fig:Comparison} and~\ref{fig:TCPU}.}
	\label{fig:sir}
\end{figure*}

Since Hyper-SIS captures the essential ingredients of algorithms for higher-order spreading processes, its extension to other dynamical models is straightforward. For instance, in the SIR epidemic process, where healed nodes remain permanently recovered, the only difference is that recovered nodes cannot be reinfected, while infection and healing rules are otherwise identical to those of SIS. Figure~\ref{fig:sir} shows the performance analysis for spreading on hypergraphs with power-law distributions of node degree and hyperedge order. The initial condition consists of a single infected node chosen at random, and simulations run until the absorbing state is reached.

The order parameter is the final fraction of recovered individuals, $\rho_{\text{R}}$. In the low-incidence regime ($\rho_{\text{R}}=0.1$), both HB-OGA and NB-OGA exhibit nearly linear computational complexity, with HB-OGA outperforming NB-OGA by a factor of nearly four. Notably, IOGA provides only a slight improvement for the node-based approach, while no noticeable difference is observed for the hyperedge-based approach; the curves for HB-OGA and HB-IOGA are indistinguishable at this scale. This occurs because high-order hyperedges are rarely activated in this regime, rendering the IOGA method effectively redundant. In contrast, in the higher-incidence regime ($\rho_{\text{R}}=0.3$), HB-OGA suffers a significant loss of efficiency, which is recovered when using IOGA, whereas NB-OGA maintains a computational complexity similar to that observed in the low-incidence regime.

Indeed, the generalization to other spreading processes requires only minor adjustments. As an example, consider a version of the Maki$-$Thompson model for rumor spreading~\cite{DeArruda2022} on higher-order networks, implemented using NB-OGA. This model features states analogous to SIR dynamics: susceptible individuals correspond to ignorants, who are unaware of the rumor; infected individuals correspond to spreaders, who propagate the rumor to ignorants; and recovered individuals correspond to stiflers, who are aware of the rumor but no longer interested in spreading it. The spreading of the rumor follows the same mechanism as infection in the SIR model, occurring at rate $\beta(m)$ for each active hyperedge of order $m$ to which a node belongs to. In contrast, the stifling (recovery) process occurs at rate $\alpha(m)$ when spreaders belong to hyperedges of order $m$ containing a number of aware individuals (spreaders plus stiflers) above a threshold $\phi(m)$; such hyperedges are referred to as saturated.

Thus, the spreading dynamics closely mirrors that of the SIR model: a list of quiescent ignorants, i.e., those belonging to at least one active hyperedge, is constructed, and one node is selected with weight $w_i \propto \boldsymbol{\beta} \cdot \mathbf{k}_i$. The selected node becomes a spreader with probability $\boldsymbol{\beta} \cdot \mathbf{n}_i / (\boldsymbol{\beta} \cdot \mathbf{k}_i)$, where $\mathbf{n}_i$ is the actual number of active hyperedges to which node $i$ belongs to; otherwise, a phantom process occurs. The stifling dynamics requires an additional list of quiescent spreaders, defined as those belonging to at least one saturated hyperedge. A node from this list is selected with probability proportional to $\boldsymbol{\alpha} \cdot \mathbf{k}_i$ and becomes a stifler with probability $\boldsymbol{\alpha} \cdot \mathbf{z}_i / (\boldsymbol{\alpha} \cdot \mathbf{k}_i)$, {where $\mathbf{z}_i$ is the actual number of saturated hyperedges to which node $i$ belongs to}; otherwise, a phantom process occurs. Both lists of quiescent ignorants and spreaders are updated using the same procedures employed for quiescent nodes in the SIR dynamics.

\subsection*{Counting rejection and phantom events}

The computational complexity of the algorithms is strongly dependent on the number of rejections during the simulation and becomes critical when the order and interaction distributions are broad, with upper cutoffs that diverge with system size. We quantify the acceptance probability during sampling for different algorithms, considering power-law order distributions with a rigid upper cutoff; see Supplementary Fig. 2. The acceptance probability decays approximately as a power law with system size, implying a higher computational cost for larger systems. On the other hand, the fraction of attempted infections that result in phantom processes is largely size-independent and remains stable across prevalence regimes. Therefore, the contribution of the dynamical evolution -- excluding the random sampling step used to select nodes and hyperedges -- has a low computational complexity of $\mathcal{O}(N)$. Consequently, if $f_{\text{acc}}(N)$ denotes the probability that a sampling attempt is accepted [see Supplementary Fig. 2(a)], the CPU time is expected to scale as $t_{\text{CPU}} \sim N / f_{\text{acc}}(N)$; see Supplementary Fig. 2(b).

\section*{Discussion}
\label{sec:Discussion}

We developed the optimized Gillespie algorithm~\cite{Cota2017}, originally introduced for pairwise spreading processes, to simulate Markovian contagion models on higher-order networks of large size and strong heterogeneity in both node degrees and hyperedge orders. Our analysis focused on the Hyper-SIS model with a critical mass threshold~\cite{deArruda2024}, enabling the assessment of algorithmic efficiency across networks ranging from $10^3$ to $10^7$ nodes. Extensions to other dynamical processes, including Hyper-SIR and rumor spreading with higher-order interactions, were also presented. Synthetic benchmarks were generated using the bipartite configuration model~\cite{Courtney}, and real-world datasets were examined~\cite{Benson2018,Landry2023}.

We proposed two algorithms based on the concept of phantom processes~\cite{Cota2017}, in which the exhaustive list of potential events required by the standard GA -- associated with high computational cost -- is replaced by a list containing both actual and phantom events that advance time without altering the system state. In the node-based OGA, the list tracks quiescent nodes belonging to active hyperedges, whereas the hyperedge-based OGA monitors potentially active hyperedges. These optimized approaches are up to several orders of magnitude faster than the standard GA, with CPU times scaling nearly linearly with system size ($t_\text{CPU} \sim N$) for bounded order and interaction distributions, compared to the quadratic scaling ($t_\text{CPU} \sim N^2$) of the standard method. For distributions with upper cutoffs increasing with system size, the computational complexity departs from strict linearity, but remains significantly more efficient than the quadratic scaling of the standard algorithm. The node-based version outperforms the hyperedge-based one in highly heterogeneous networks, whereas the reverse holds for low-heterogeneity structures. In addition, the adoption of efficient rejection schemes to sample nodes or hyperedges further improves performance, especially in strongly heterogeneous systems.

In the low-prevalence regime, where contagion is dominated by pairwise interactions, our results agree with previous studies on purely pairwise contagion models~\cite{Cota2017,Costa2021,St-Onge2019}. In contrast, in high-prevalence regimes, where higher-order interactions sustain the spreading dynamics, the proposed algorithms exhibit substantially lower computational complexity than the standard GA. While GA scales as $t_\text{CPU} \sim N^2$, HB-OGA and NB-OGA achieve nearly linear scaling with system size, representing the lower computational bound in non-vanishing prevalence regimes. The relative performance of the two optimized methods depends on the strength of higher-order effects: as these interactions become stronger, the node-based approach (NB-OGA) increasingly outperforms the hyperedge-based one (HB-OGA). The observed $t_{\text{CPU}} \sim N^2$ scaling of the standard algorithm arises from the finite-prevalence regime ($\rho = 0.1-0.8$) considered here, where a macroscopic fraction of higher-order interactions remains active, and is therefore not directly tied to structural density. In contrast, the optimized algorithms approach the minimal achievable complexity $t_{\text{CPU}} \sim N$ under bounded interaction sizes, highlighting their ability to efficiently handle configuration-dependent multi-body event spaces.

Finally, we also analyzed simplicial complexes, a specific class of higher-order networks~\cite{Torres, Bianconi2021}. These structures are densely connected at lower orders, resulting in a large number of links per node, and are frequently used as a framework for studying contagion dynamics~\cite{Iacopini,wang2021simplicial}. Their performance characteristics are consistent with those observed in Fig.~\ref{fig:CPU_Hom} for networks with homogeneous interaction sizes and low order heterogeneity: HB-OGA and NB-OGA achieve comparable performance with low computational complexity, scaling as $t_\text{CPU} \sim N$, and both significantly outperform the standard GA, which exhibits the much higher complexity $t_\text{CPU} \sim N^2$.

Although initially motivated by SIS-type contagion dynamics and extended here to SIR epidemic models and Maki--Thompson rumor spreading dynamics, the algorithms can be readily generalized to other processes on hypergraphs exhibiting inactive-to-active transitions, including rumor propagation~\cite{DeArruda2022,Daley,oliveira2026rumor} and other epidemic models such as SIRS and SEIR~\cite{Pastor-Satorras2015,zhang2023dynamical,wang2021simplicial}. More broadly, the relevance of the proposed approach stems from the fact that higher-order dynamical processes are governed by intrinsically multi-body interaction rules, for which the set of feasible transitions is configuration-dependent and cannot be efficiently handled within standard pairwise frameworks. In this context, the algorithms introduced here provide a scalable and statistically exact methodology for simulating such systems. We therefore expect them to become versatile tools for investigating spreading and activation dynamics in large, heterogeneous higher-order networks, enabling the systematic exploration of models and phenomena that remain inaccessible to conventional simulation approaches.

\section*{Methods}

\subsection*{The optimized Gillespie algorithm for SIS models with pairwise interactions}
\label{sec:GA}

Before discussing epidemic models, consider a set of $Z$ independent Poisson processes, $p = 1, 2, 3, \dots, Z$, each occurring at rate $\lambda_p$, such that the probability that an event $p$ occurs within the time interval $[t, t+dt]$ is $\lambda_p dt$. The total rate at which any event occurs is $R = \sum_{p=1}^Z \lambda_p$, implying that the probability density function for an event to occur at time $t$ is
\begin{equation}
	\pi(\tau) = R \left( \prod_{p=1}^{Z} e^{-\lambda_p \tau} \right) = Re^{-R \tau}.
	\label{eq:Poisson}
\end{equation}
The Gillespie algorithm (GA)~\cite{Gillespie} is a statistically exact method for simulating Markovian stochastic reaction processes. It proceeds through discrete time steps $\tau$ drawn from an exponential distribution as defined in Eq.~\eqref{eq:Poisson}, and at each step selects one event with probability proportional to its rate $\lambda_p$. The list of all possible events is continuously updated during the simulation.

The standard (non-optimized) GA applied to the SIS model on pairwise networks involves two lists of processes: $\Lambda^{(\text{I})}$, containing all $N_\text{I}$ infected nodes, and $\Lambda^{(\text{IS})}$, containing all $N_\text{IS}$ edges connecting infected to susceptible nodes. The total rate has two contributions, $R = F + G$, where
\begin{equation}
	F = \sum_{i=1}^N \alpha \sigma_i =  \alpha N_\text{I},
	\label{eq:HealRate1}
\end{equation}
is the total healing rate, and
\begin{equation}
	G = \sum_{i,j} A_{ij}  \beta \sigma_i (1 - \sigma_j) =  \beta N_\text{IS},
	\label{eq:PWInfectRate}
\end{equation}
is the total infection rate, with $A_{ij} = {A}_{\{i,j\}}^{(1)}$ being the standard adjacency matrix~\cite{Barabasi}. At each time step, with probability $f = F/R$, one node is randomly selected from the list $\Lambda^{(\text{I})}$ and healed. With complementary probability $1-f = G/R$, one IS edge is randomly chosen from the list $\Lambda^{(\text{IS})}$, and the corresponding susceptible node is infected. Time is incremented by $\tau = -\ln u / R$, where $u$ is a pseudo-random number uniformly distributed in the range $(0,1)$, and both lists are updated.

Updating the infection lists has high computational complexity, which quickly becomes prohibitive even for networks of only tens of thousands of nodes. This limitation can be largely overcome by aforementioned phantom processes~\cite{Cota2017}. In epidemic models, this corresponds to overcounting infection attempts by assuming that each infected node transmits to all of its neighbors at rate $\beta$, regardless of their state, while an actual infection occurs only if the selected neighbor is susceptible. This leads to a total attempt infection rate
\begin{equation}
	G = \sum_{i} \beta k_i \sigma_i,
	\label{eq:PWInfectRate2}
\end{equation}
which is greater than or equal to the true infection rate in Eq.~\eqref{eq:PWInfectRate}. The rest of the algorithm proceeds analogously. The main different is that when an infection attempt is chosen, an infected node from the list $\Lambda^{(\text{I})}$ (no longer requiring $\Lambda^{(\text{IS})}$) is selected with probability proportional to its degree $k_i$. One of its neighbors $j$ is then chosen at random: if $j$ is susceptible, it becomes infected; otherwise, a phantom process occurs and the system state remains unchanged. In both cases, time is advanced by $\tau = -\ln u / R$, with $R$ computed from Eqs.~(\ref{eq:HealRate1}) and(~\ref{eq:PWInfectRate2}). This optimization preserves the statistical exactness of the Gillespie procedure~\cite{Cota2017}.

\subsection*{Data availability }
Hypergraph datasets generated in this paper are publicly at \url{https://doi.org/10.5281/zenodo.17187745}, ref.~\cite{datasets2025}.

\subsection*{Code availability}
The HB-OGA and NB-OGA codes are available at \url{https://github.com/gisc-ufv/hyperSIS}, ref.~\cite{codes2025}. The code is developed in Modern Fortran~\cite{Curcic2020}, it follows a modular, object-oriented structure and is compatible with the Fortran Package Manager (\texttt{fpm})~\cite{Kedward2022}, as well as a Python interface. A Jupyter Notebook with usage examples is provided. Network input can be supplied as a list of hyperedges, in bipartite format, in the XGI JSON format~\cite{Landry2023}, or in the HIF standard format~\cite{coll2025hif}. Both temporal and quasi-stationary dynamics are available. The code was run using the LLVM-based Intel Fortran (\texttt{ifx}) and the non-commercial GNU Fortran (\texttt{gfortran}) compilers, on Linux and Windows Subsystem for Linux (WSL). 

\subsection*{Acknowledgments}
We thank M. Clarin from COSNET Lab-BIFI for helping with the figures. 

\subsection*{Funding}
H.P.M, S.C.F, and W.C. acknowledge the financial support of Fundação de Amparo à Pesquisa do Estado de Minas Gerais (FAPEMIG)-Brazil (Grants No. APQ-01973-24 and APQ-03079-24). S.C.F. acknowledges the financial support by the Conselho Nacional de Desenvolvimento Científico e Tecnológico (CNPq)-Brazil (Grants no. 310984/2023-8 and 408389/2024-9 INCT-NeuroComp) and Fundação de Amparo à Pesquisa do Estado de São Paulo (FAPESP)-Brazil (Grant. No. 25/24366-1). W. C. acknowledges the financial support by INCT-DigiSaúde (CNPq Grant 408775/2024-6). Y. M. was partially supported by the Government of Aragon, Spain, and ERDF "A way of making Europe" through grant E36-23R (FENOL). YM also acknowledges support from Grant No. PID2023-149409NB-I00 from Ministerio de Ciencia, Innovación y Universidades, Agencia Española de Investigación (MICIU/AEI/10.13039/501100011033) and ERDF ``A way of making Europe''. This study was financed in part by the Coordenação de Aperfeiçoamento de Pessoal de Nível Superior (CAPES), Brazil, Finance Code 001.

\subsection*{Author Contributions}
H.P.M: Conceptualization, formal analysis, methodology, data curation, coding and simulations, visualization, writing original draft. W.C.: Validation, methodology, data curation, formal analysis, coding and simulations,  writing and editing. Y.M.: Conceptualization, formal analysis, writing and editing, and funding acquisition. S.C.: Conceiving the study, conceptualization, formal analysis, methodology, writing original draft, supervision, and funding acquisition.

\bibliography{ref1}

\end{document}


\title{Supplementary Information: Efficient Gillespie algorithms for spreading phenomena in large and heterogeneous higher-order networks}

\author{Hugo P. Maia}
\affiliation{Departamento de F\'{\i}sica, Universidade Federal de Vi\c{c}osa, 36570-900, Vi\c{c}osa, MG, Brazil}

\author{Wesley Cota}
\affiliation{Departamento de F\'{\i}sica, Universidade Federal de Vi\c{c}osa, 36570-900, Vi\c{c}osa, MG, Brazil}

\author{Yamir Moreno}
\affiliation{Institute for Biocomputation and Physics of Complex Systems (BIFI), University of Zaragoza, Zaragoza 50009, Spain}
\affiliation{Department of Theoretical Physics, Faculty of Sciences, University of Zaragoza, Zaragoza, Spain}

\author{Silvio C. Ferreira}
\affiliation{Departamento de F\'{\i}sica, Universidade Federal de Vi\c{c}osa, 36570-900, Vi\c{c}osa, MG, Brazil}
\affiliation{National Institute of Science and Technology for Complex Systems, Centro Brasileiro de Pesquisas F\'{\i}sicas, Rua Xavier Sigaud 150, 22290-180, Rio de Janeiro, Brazil}

\date{\today}

\maketitle

\section*{Supplementary Methods: Construction of synthetic higher-order networks}
\label{sec: BCM}

\begin{figure*}[hbt]
	\centering
	\includegraphics[width=0.8\linewidth]{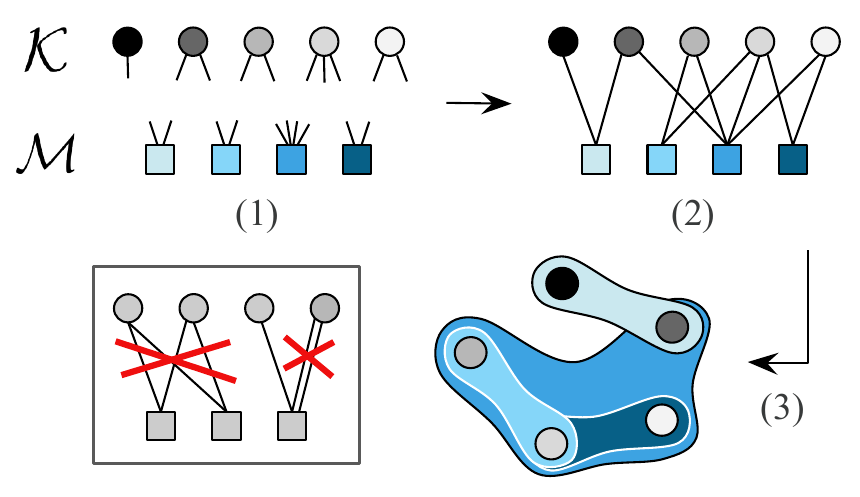}
	\caption{\justifying \textbf{Schematic representation of bipartite configuration model (BCM).}
		(1) A bipartite graph is built with five nodes in partition $\mathcal{K}$ (circles) and four in partition $\mathcal{M}$ (squares), whose degrees are drawn from predefined distributions $P_K$ and $f_m$.
		(2) Links are created only between nodes belonging to different partitions.
		(3) The resulting bipartite graph is interpreted as a higher-order network with five nodes and four hyperedges; repeated nodes and duplicate hyperedges are excluded.}
	\label{fig:BipartiteCM}
\end{figure*}

Synthetic hypergraphs were generated using the adapted bipartite configuration model~\cite{Courtney}. The BCM independently prescribes (i) the distribution of interactions per node and (ii) the distribution of group sizes, and then couples them through a bipartite matching process. This separation allows us to tune connectivity heterogeneity and higher-order heterogeneity in a controlled and reproducible way. A bipartite graph is built from two disjoint node sets, $\mathcal{K}$ and $\mathcal{M}$, with degree sequences $\{k_i\}$ and $\{m_j\}$ drawn from distributions, $P_K$ and $f_m$, respectively. Stubs from opposite partitions are paired uniformly at random to form the bipartite network. The resulting structure is then projected into a higher-order network by interpreting nodes in $\mathcal{K}$ as agents and each node $j\in\mathcal{M}$ as a hyperedge containing its $m_j$ neighbors. In this mapping, a hyperedge of order $m$ corresponds to a node in $\mathcal{M}$ with degree $m\!+\!1$ (see Fig.~\ref{fig:BipartiteCM}).

To avoid artifacts, we forbid duplicate hyperedges and repeated nodes within the same hyperedge. The most expensive step is checking potential duplicates among hyperedges of the same order. In practice, this can be handled efficiently with hashing and sorting techniques~\cite{Batagelj2005}, which allow efficient detection and removal of duplicates during generation. Moreover, for a broad range of degree and order distributions, the frequency of duplicates is sufficiently low to be negligible in large systems; see Ref.~\cite{Courtney} for details.

\pagebreak

\section*{Supplementary Figures}
\label{sec: Figures}

\begin{figure*}[h]
	\centering
	\begin{subfigure}[b]{0.425\textwidth}
		\centering
		\includegraphics[width=\linewidth]{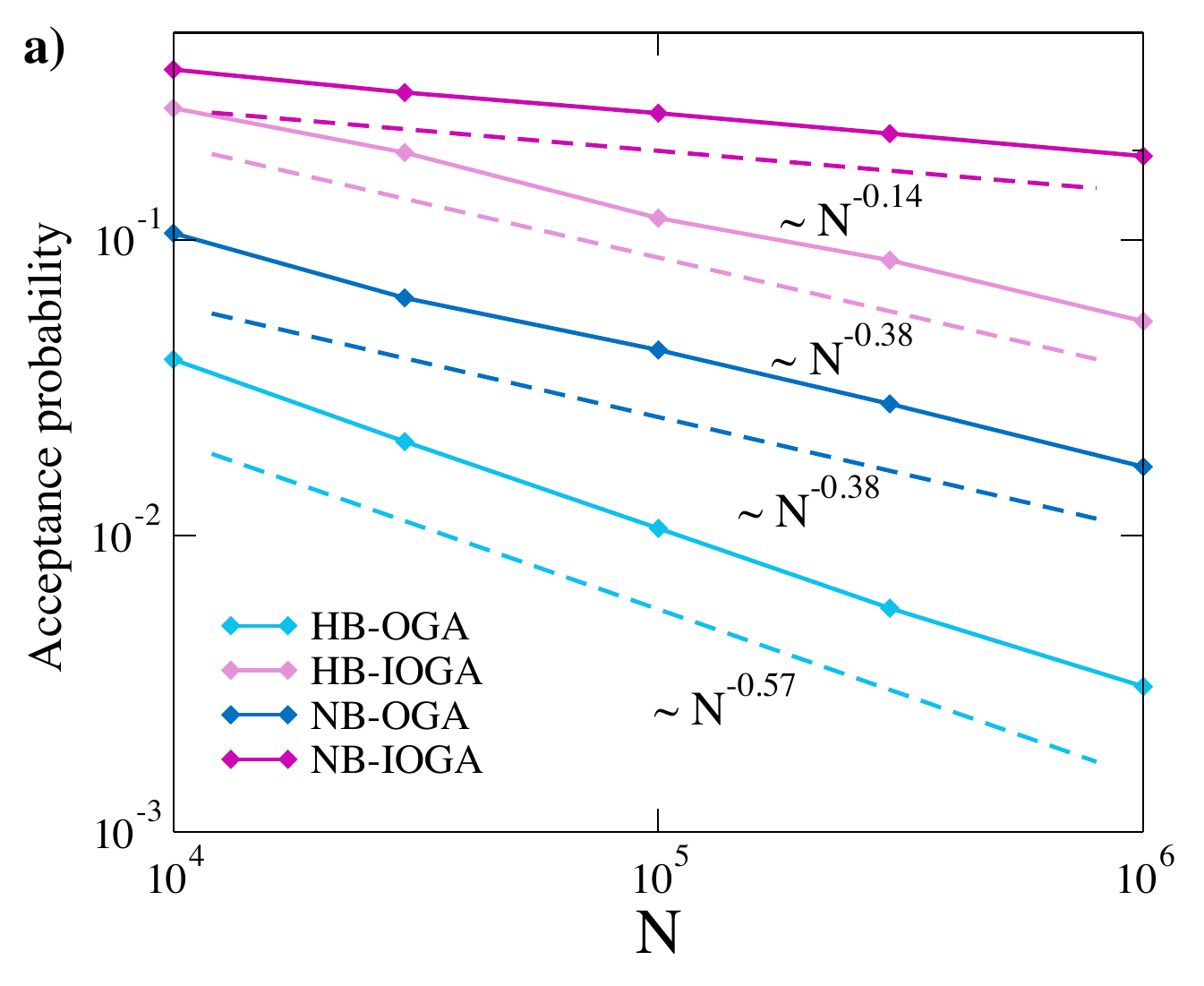}
	\end{subfigure}
	\begin{subfigure}[b]{0.425\textwidth}
		\centering
		\includegraphics[width=\linewidth]{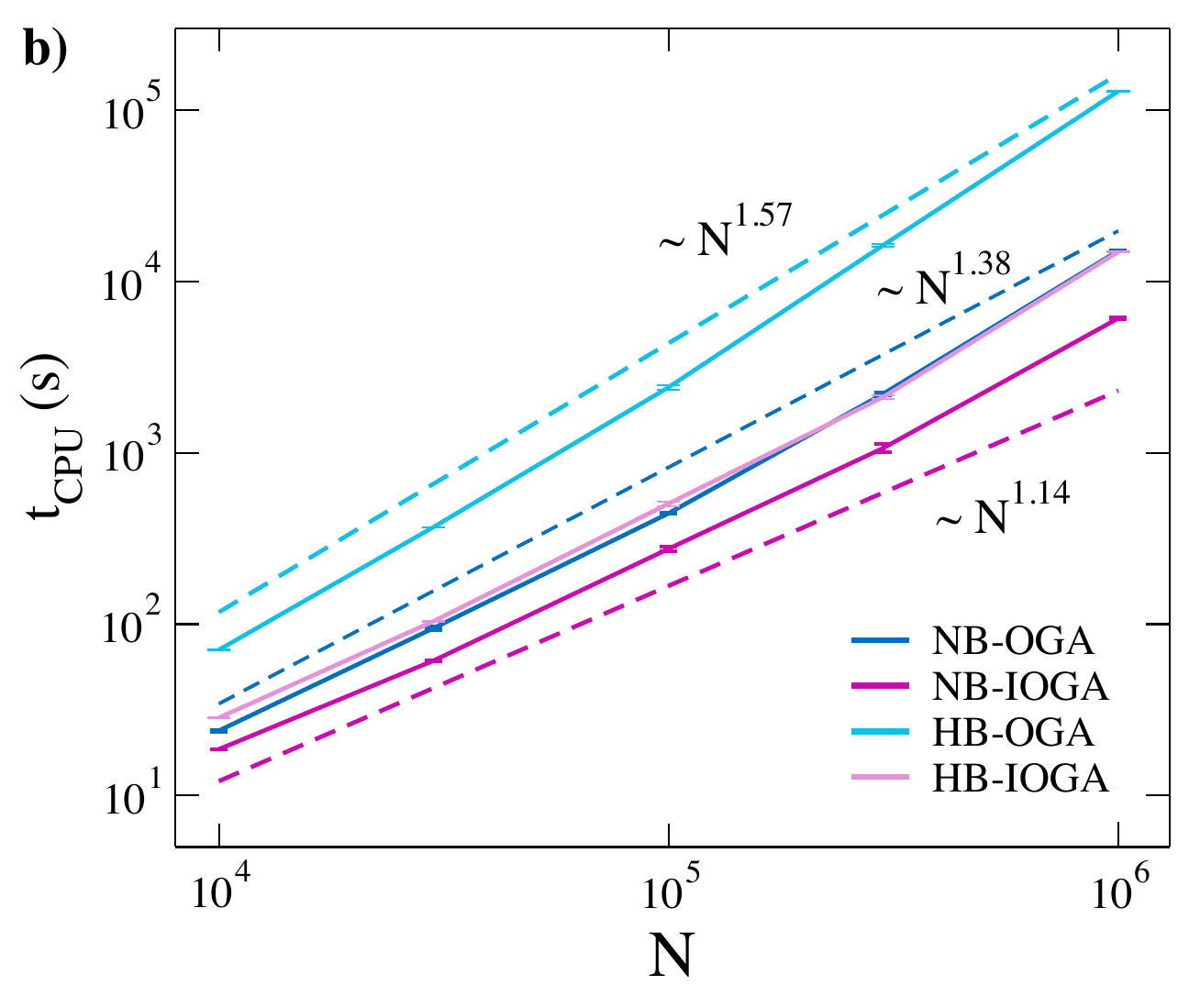}
	\end{subfigure}
	\caption{\justifying \textbf{Analysis of rejection for different algorithms}  (a) Probability of acceptance during the sampling to select node or hyperdges for different algorithms.  (b) CPU times in seconds validating the hypothesis that $t_\text{CPU}\sim N/f_\text{ac}(N)$. The simulations were run on  a power-law higher-order network with degree and order exponent set to $\gamma_k=3$ and $\gamma_m = 3$, respectively.
	}
	\label{fig:Rej}
\end{figure*}

\begin{figure*}[h]
	\begin{center}
		\includegraphics[width=0.7\textwidth]{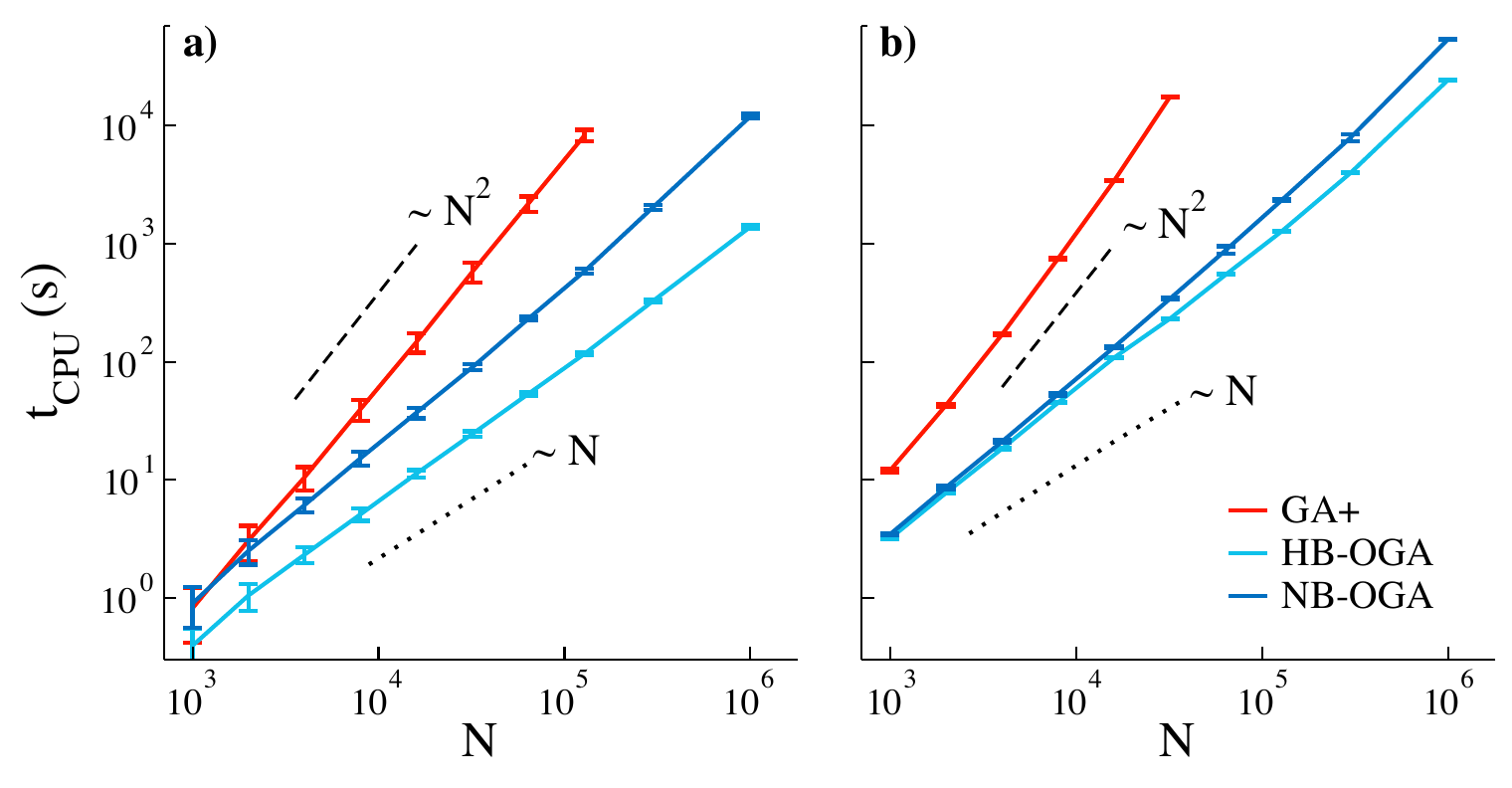}
	\end{center}
	\caption{\justifying \textbf{Algorithms' performance for different epidemic prevalences.} CPU times for simulation of Hyper-SIS contagion dynamics on hypergraphs with power-law order distribution ($\gamma_m = 3.0$) using different algorithms. Simulations were performed with $b = 0.4$ and $\theta_0 = 1.0$, with the first-order spreading rate $\beta(1)$ set to maintain a stationary infection density of (a) $\rho \approx 0.1$ and (b) $\rho \approx 0.7$ for all sizes. {Heterogeneous interaction distribution was considered for fixed $\gamma_k = 2.7$ and the degree range $K \in [3, K_\text{c}]$, where $K_\text{c}(N)$ is a rigid cutoff.}}
	\label{fig:CPU_rho}
\end{figure*}

\bibliography{ref1}